\documentstyle[12pt,prc,aps]{revtex}
\begin{document}
\draft

\title{
Spin-Isospin Response Functions \\
and
the Effects of the $\Delta$-Hole
Configurations \\
in Finite Nuclei
}

\author{
Kimiaki NISHIDA\footnote{
Electronic address: knishida@hep1.c.u-tokyo.ac.jp}
and
Munetake ICHIMURA\footnote{
Electronic address: ichimura@tansei.cc.u-tokyo.ac.jp}
}

\address{
Institute of Physics, University of Tokyo, Komaba \\
Komaba, Meguro-ku, Tokyo 153, Japan
}

\date{\today}

\hfill UT-Komaba 94-19

\hfill nucl-th/9410003

\begin{center}
{\Large \bf
Spin-Isospin Response Functions \\
and
the Effects of the $\Delta$-Hole
Configurations \\
in Finite Nuclei \\
}

\vspace{6mm}
Kimiaki NISHIDA\footnote{
Electronic address: knishida@hep1.c.u-tokyo.ac.jp}
and
Munetake ICHIMURA\footnote{
Electronic address: ichimura@tansei.cc.u-tokyo.ac.jp}

\vspace{6mm}
{\it
Institute of Physics, University of Tokyo, Komaba \\
Komaba, Meguro-ku, Tokyo 153, Japan
}
\end{center}

\vspace{6mm}
\begin{abstract}
\setlength{\baselineskip} {6mm}
Effects of the delta-isobar $(\Delta)$ mixing
on the spin-isospin response functions in finite nuclei
are studied in the quasi-elastic region.

A method to calculate the response function
for a finite system composed of nucleon $(N)$ and $\Delta$
is formulated in a ring approximation.
It is designed to treat
the $\Delta$-related Landau-Migdal parameters,
$g'_{N\Delta}$ and $g'_{\Delta\Delta}$, and
the nucleon parameter $g'_{NN}$, independently,
so that the universality ansatz,
$g'_{NN}=g'_{N\Delta}=g'_{\Delta\Delta}$,
is removed.

We calculated the isovector spin-longitudinal and
-transverse response functions, $R^{}_{L}$ and $R^{}_{T}$,
with and without the $\Delta$-mixing.
Inclusion of $\Delta$ enhances $R^{}_{L}$
but reduces $R^{}_{T}$
for ordinary interactions.
Dependence of $R^{}_{L,T}$ on $g'_{NN}$ and $g'_{N\Delta}$
is investigated.
Decomposition into the process-decomposed response
functions,
$R^{[NN]}_{}$,
$R^{[N\Delta]}_{}$ and
$R^{[\Delta\Delta]}_{}$,
is very elucidative to see the $\Delta$ effects,
which are found to be mainly governed by
$R^{[N\Delta]}_{}$ and sensitive to $g'_{N\Delta}$.

The isovector spin-transverse response function
$R^{(e,e')}_{T}$
obtained by $(e,e')$
is calculated by various effective interactions
and compared to each other as well as experimental data.
\end{abstract}

\setlength{\baselineskip} {6mm}

\section{Introduction}
\label{sec:Introduction}

The spin-isospin properties of nuclei are a long standing
and
still very interesting subject
of nuclear physics
\cite{Ichimura91,Osterfeld92}.
New theoretical and experimental developments are seen
in the study of nuclear spin-isospin response functions
over the last decade.
They promoted many interesting problems
and active researches in this fields
\cite{Osterfeld92}.

In the analysis of the responses in the quasi-elastic
region,
nuclear currents with the spin and isospin degrees of
freedom
are usually separated into
the isovector spin-longitudinal and -transverse components.
They are characterized by the operators,
$\mbox{\boldmath ${\tau}$} (\mbox{\boldmath ${\sigma}$}
\cdot \widehat{\mbox{\boldmath ${q}$}})$ and
$\mbox{\boldmath ${\tau}$} (\mbox{\boldmath ${\sigma}$}
\times \widehat{\mbox{\boldmath ${q}$}})$,
respectively,
where $\mbox{\boldmath ${q}$}$ is the transferred momentum
to the nucleus
and $\mbox{$ \widehat{\mbox{\boldmath $q$}}$} \equiv
\mbox{\boldmath $q$} / q$.
The spin-transverse response function $R^{}_{T}(q,\omega)$
has long been observed
in a wide range of $q$ and the transferred energy $\omega$
by electron scattering
\cite{Altemus80,Barreau83,Deady83,Meziani85,Blatchley86}.
However, hadronic probes are needed
to study the spin-longitudinal response
function $R^{}_{L}(q,\omega)$.

Owing to the great progress of the experimental technique,
complete measurement of the polarization transfers $D_{ij}$
was first carried out at LAMPF
\cite{Carey84,Rees86}
for the quasi-elastic $(\vec{p},\vec{p}')$ scattering,
from which they extracted the ratio $R^{}_{L}/R^{}_{T}$
by use of an eikonal approximation.
Measurement of $(\vec{p},\vec{n})$ reaction is more
difficult
but more preferable
because it can exclusively extract the isovector part.
Recently LAMPF group
\cite{McClelland92,Chen93,Taddeucci94}
observed $D_{ij}$ of
the quasi-elastic ($\vec{p}$,$\vec{n}$) reaction
and extracted $R^{}_{L}/R^{}_{T}$ in a similar way.
An interesting finding of these experiments is
that the ratio is close to or less than unity
in the quasi-elastic region.

This was surprising because the standard nuclear model
predicted
that the ratio was much larger than unity
at relatively low $\omega$.
Based on the random-phase approximation (RPA) in nuclear
matter,
Alberico {\it et al.}
\cite{Alberico82}
pointed out
that $R^{}_{L}$ is enhanced
and its peak position is shifted downwards (softening),
whereas $R^{}_{T}$ is quenched
and its peak is shifted upwards (hardening)
around the transferred momentum $q=1.75 \;{\rm fm}^{-1}$.
Therefore the ratio $R^{}_{L}/R^{}_{T}$
extremely exceeds unity
especially at relatively low $\omega$.
However, such behavior has hardly been seen in the
experiments.

This contradiction
has been challenged from various aspects,
such as finite size effect of nucleus
\cite{Alberico85,Alberico86,Shigehara88,Ichimura89},
relativistic RPA approach
\cite{Horowitz93},
nuclear correlations beyond RPA
\cite{Alberico84,Smith88,Takayanagi90,Takayanagi93,Pandharipande94},
effects of absorptions and distortions
\cite{Ichimura89,Izumoto82,Esbensen85B,Smith88B,Sams93},
etc..
Among them
here we investigate
effects of $\Delta$-mixing
and
dependence on the nucleon particle-hole ($ph$)
and $\Delta$-hole ($\Delta h$) effective interactions.

The analysis of Alberico {\it et al.}
\cite{Alberico82}
took account of the $\Delta$ degree of freedom and
used the ($\pi + \rho + g'_{}$) model
for the effective interaction,
namely
one-pion exchange +
one-rho-meson exchange +
the contact interaction
specified by the Landau-Migdal parameters $g'_{}$'s.
There appear three different $g'$'s relevant to the
interactions
between $ph$ and $ph$, between $ph$ and $\Delta h$ and
between $\Delta h$ and $\Delta h$.
They are denoted by $g'_{NN}$, $g'_{N\Delta}$
and $g'_{\Delta\Delta}$, respectively.

For computational simplicity,
most of previous works adopted
the universality ansatz
\cite{Oset79,Oset82},
$g'_{NN} = g'_{N\Delta} = g'_{\Delta\Delta}$.
However it has no theoretical foundation
nor was supported by various estimations
\cite{Osterfeld92}.
For instance, the phenomenological analysis
\cite{Meyer-ter-Vehn81,Toki81}
yields $g'_{NN} \approx 0.6 \sim 0.7$.
$G$-matrix calculations
suggested
$g'_{NN} \approx 0.5$ and $g'_{N\Delta} \approx 0.4$,
but the induced interaction increases and
the finite size effect decreases them
\cite{Dickhoff81,Dickhoff83,Cheon84,Nakayama84,Czerski86,Dickhoff87}.
A recent estimation by Brown {\it et al.}
\cite{Brown93}
gives
$g'_{NN} \approx 1.0$,
$g'_{N\Delta} \approx 0.33$
and $ g'_{\Delta\Delta} \approx 0.5$.

Considering this situation, we developed a RPA formalism
for the finite nucleus composed of $N$ and $\Delta$,
in which the universality ansatz is not adopted.
Then we investigated
the effects of $\Delta$
and
the $g'$ dependence of $R^{}_{L}$ and $R^{}_{T}$.
Similar analysis was performed in the Fermi gas model
by Shiino {\it et al.}
\cite{Shiino86}.
However the model gives
unreasonably large enhancement and softening of $R^{}_{L}$
\cite{Ichimura89}
and hence it is somewhat misleading.
Here we perform a more realistic and detailed analysis
for {\it finite} nuclei.

The response functions including $N$ and $\Delta$
constitute of three different components,
$R^{[NN]}_{}$,
$R^{[N\Delta]}_{} ( = R^{[\Delta N]}_{}$) and
$R^{[\Delta\Delta]}_{}$,
which correspond to the processes depicted in
Fig.~\ref{fig:WDIAG_A}.
$R^{[NN]}_{}$ represents the process in which
the nuclear current first creates a $ph$ state
and annihilates it at the end.
$R^{[N\Delta]}_{}$ corresponds to the process which start
with
the $\Delta h$ creation
and end with the $ph$ annihilation, and
$R^{[\Delta N]}_{}$ corresponds to the inverse one.
$R^{[\Delta\Delta]}_{}$ represents those which start from
the $\Delta h$ excitation
and end with its annihilation.
We call them the process-decomposed response functions.

We will see that
the $\Delta h$ configuration is crucial for
the enhancement of $R^{}_{L}$
and
plays some role for the quenching of $R^{}_{T}$
for ordinary interactions.
These effects come mainly from $R^{[N\Delta]}_{}$,
which is sensitive to $g'_{N\Delta}$.
This indicates that we must avoid the universality ansatz
and treat $g'$'s independently.

In many of the previous works with the universality ansatz,
the decomposition did not appear explicitly.
The response functions have implicitly been calculated by
\begin{eqnarray}
R^{}_{L,T} = R^{[NN]}_{L,T}
   + 2 \frac{f_{\Delta}}{f_{N}}
     R^{[N\Delta]}_{L,T}
   + \left(\frac{f_{\Delta}}{f_{N}}\right)^{2}
     R^{[\Delta\Delta]}_{L,T}
\label{DefRLRT}
\label{eq:NA}
\end{eqnarray}
where $f_{N}$ and $f_{\Delta}$ are
the $\pi NN$ and $\pi N\Delta$ coupling constants,
respectively,
and $f_{\Delta} = 2.0 f_{N}$ is usually used.
However, observed response functions depend on probes.
For instance,
the isovector spin-transverse response function
$R^{(e,e')}_{T}$
observed by $(e,e')$
is approximated by
\begin{eqnarray}
R^{(e,e')}_{T} = R^{[NN]}_{T}
  + 2 \frac{f_{\gamma N\Delta}}{f^{IV}_{\gamma NN}}
    R^{[N\Delta]}_{T}
  + \left(\frac{f_{\gamma N\Delta}}{f^{IV}_{\gamma
NN}}\right)^{2}
R^{[\Delta\Delta]}_{T}
\end{eqnarray}
where $f^{IV}_{\gamma NN}$ and $f_{\gamma N\Delta}$ specify
the isovector magnetic coupling strength of
$\gamma NN$ and $\gamma N\Delta$ vertices, respectively.
We also check this coupling constant dependence.

In Sec.~\ref{sec:Formalism}
we present a formalism for calculating the response
functions
under the ring approximation
in the finite system
which is composed of $N$ and $\Delta$.
The finiteness is handled by the continuum RPA
with the orthogonality condition
\cite{Izumoto83,Ichimura89,Kawahigashi91}.
The universality ansatz for $g'$ is removed.
In Sec.~\ref{sec:DeltaHole}
we show
the effects of the $\Delta$ mixing on the response
functions,
and
analyze them in the form of the process-decomposed
components.
The energy-weighted and non-weighted sums are also
discussed.
In Sec.~\ref{sec:EffectivInt}
we investigate
the effective interaction dependence of
$R^{}_{L,T}$ and
$R^{[\alpha\beta]}_{L,T}$.
We compare the results obtained by various values of
$g'_{}$.
We also show some results with
lighter $\rho$-meson effective mass $m^{\ast}_{\rho}$.
In Sec.~\ref{sec:ElectronScattering}
we present our calculation of $R^{(e,e')}_{T}$ with
experimental data.
Its dependence on $g'_{}$ and
the ratio $f_{\gamma N\Delta}/f^{IV}_{\gamma NN}$
is shown.
The summary is given in Sec.~\ref{sec:Summary}.

\section{Formalism}
\label{sec:Formalism}

In this section,
we formulate a method to calculate the response functions
for a finite system composed of $N$ and $\Delta$
in the ring approximation.
For simplicity,
we only consider doubly (sub-)closed shell nuclei,
so that
the particle and the hole states are well defined
and
the spin of
the ground-state is zero.

We express
the spin and the isospin operators of $N$
$({\mbox{\boldmath $\sigma$}}^{}_{}$ and ${\mbox{\boldmath
$\tau$}}^{}_{})$
and
the transition operators
between $N$ and $\Delta$
$(\mbox{\boldmath $S$}$ and $\mbox{\boldmath $T$})$
in unified spherical tensor forms,
${\sigma}^{s(ab)}_{\mu}$ and
${\tau}^{t(ab)}_{\nu}$,
as
\begin{eqnarray}
\begin{array}{ll}
\hskip 7.15em
{\sigma}^{0(ab)}_{0}
  \equiv \delta_{ab} & \hskip 4em \mbox{for}\;\; s=0, \\
{\sigma}^{1(NN)}_{\mu}
  \equiv {\sigma}^{}_{\mu}, \;\;
{\sigma}^{1(\Delta N)}_{\mu}
  \equiv S_{\mu}, \;\;
{\sigma}^{1(N \Delta)}_{\mu}
  \equiv (-)^{\mu}S_{-\mu}{}^{\dagger} & \hskip 4em
\mbox{for}\;\; s=1,
\end{array}
\end{eqnarray}
with
$a,b = N$ or $\Delta$,
and in a similar way for ${\tau}^{t(ab)}_{\nu}$.

\subsection{Spin-isospin polarization propagator}
We define
the spin-isospin (transition) current operators as
\begin{eqnarray}
   j^{t\nu(ab)}_{s\mu}(\mbox{\boldmath $r$})
   &\equiv&
       \sum^{A}_{k=1}
       j^{t\nu(ab)}_{s\mu}(\mbox{\boldmath
$r$};\mbox{\boldmath $r$}_{k})
       \label{eq:AL}
       ,\\
   j^{t\nu(ab)}_{s\mu}(\mbox{\boldmath $r$};\mbox{\boldmath
$r$}_{k})
   &\equiv&
       \left[
       \left(\tau^{t(ab)}_{\nu}\right)_{k}
       \left(\sigma^{s(ab)}_{\mu}\right)_{k}
       \right]
       \delta^{3}(\mbox{\boldmath $r$}-\mbox{\boldmath
$r$}_{k})
       ,
\end{eqnarray}
where
$\mbox{\boldmath $r$}_{k}$ is
the position vector of the $k$-th particle in the nucleus
and
$j^{(ab)}_{}$ operates only on the type $(b)$ particle.
They are separated into the angle and radial parts of
$\mbox{\boldmath $r$}$ as
\begin{eqnarray}
   j^{t\nu(ab)}_{s\mu}(\mbox{\boldmath $r$})
   &=&
       \sum_{\ell JM}
       \sum_{m}
       \left[
       \langle\,\ell m s \mu\,|\,JM\,\rangle i^{\ell}
       Y^{\ell}_{m}(\Omega_{r})
       \right]^{\ast}
       j^{t\nu(ab)}_{\ell sJM}(r)
       ,
\end{eqnarray}
where
\begin{eqnarray}
   j^{t\nu(ab)}_{\ell sJM}(r)
   &\equiv&
       \sum^{A}_{k=1}
       j^{t\nu(ab)}_{\ell sJM}(r;\mbox{\boldmath $r$}_{k})
       \label{eq:AN}
       , \\
   j^{t\nu(ab)}_{\ell s JM}(r;\mbox{\boldmath $r$}_{k})
      &\equiv&  \frac{\delta(r-r_{k})}{rr_{k}}
                \left(\tau^{t(ab)}_{\nu}\right)_{k}
                \left[
                i^{\ell} Y^{\ell}_{}(\Omega_{r_{k}})
                \otimes
                \left(\sigma^{s(ab)}_{}\right)_{k}
                \right]^{J}_{M}
      .\label{eq:ZA}
\end{eqnarray}

We then introduce
the spin-isospin polarization propagator
\cite{Alberico85}
as
\begin{eqnarray}
\Pi^{t\nu,t'\nu' (ab,cd)}_{\ell sJM, \ell'
s'J'M'}(r,r';\omega)
     &=&
           \langle\, \Psi_{0} \, |\,
          \left[
           \mbox{$ \widetilde{j}$}^{t\nu(ab)}_{\ell sJM}
           (r)
           \frac{1}{\omega - (H - E_{0}) + i\eta}
           {\mbox{$ \widetilde{j}$}^{t'\nu'(dc)}_{\ell'
s'J'M'}}{}^{\dagger}(r')
           \right. \nonumber \\ && \hskip 2em \left.
         -
           {\mbox{$ \widetilde{j}$}^{t'\nu'(dc)}_{\ell'
s'J'M'}}{}^{\dagger}(r')
           \frac{1}{\omega + (H - E_{0}) - i\eta}
           \mbox{$ \widetilde{j}$}^{t\nu(ab)}_{\ell sJM}(r)
\right]
           \,|\, \Psi_{0} \, \rangle
           \label{eq:ZB}
           \\
      &=&
            \delta_{JJ'}
            \delta_{MM'}
            \delta_{\nu\nu'}
            \Pi^{tt'\nu(ab,cd)}_{J\ell s \ell'
s'}(r,r';\omega)
           ,
\end{eqnarray}
where
$
\mbox{$ \widetilde{j}$}^{}_{}
   \equiv
   j^{}_{}
   -
   \langle\, \Psi_{0} \, |\, j^{}_{} \,|\, \Psi_{0} \,
\rangle
$
is the current fluctuation and
$\,|\, \Psi_{0} \, \rangle$ and $E_{0}$ denote the
ground-state of the nuclei
and its energy, respectively.
The second identity
comes from the assumption that
the total angular momentum of
the ground-state
is zero.
In the pure shell model without the residual interaction,
the total Hamiltonian $H$ is replaced
by the uncorrelated one $H^{(0)}$,
the sum of the single-particle Hamiltonian.
Then
the uncorrelated polarization propagator is given by
\begin{eqnarray}
{\Pi^{(0)}}^{tt'\nu(ab,cd)}_{J\ell s \ell' s'}(r,r';\omega)
     &=&
         \delta_{ad} \delta_{bc}
         \sum_{p(\Delta)h}
         \left[
         \delta_{aN}
         \frac{
                 \langle\, \Phi_{0} \, |\,
                 j^{t\nu(ab)}_{\ell sJM}(r)
                 \,|\, p(\Delta)h \, \rangle
                 \langle\, p(\Delta)h \, |\,
                 {j^{t'\nu(dc)}_{\ell'
s'JM}}{}^{\dagger}(r')
                 \,|\, \Phi_{0} \, \rangle
                }
                {\omega -
\left({\epsilon}^{(b)}_{p(\Delta)}
                        - {\epsilon}^{}_{h}\right) + i\eta}
           \right. \nonumber \\
     &&\hskip 4.5em
         \left.
         -
         \delta_{bN}
         \frac{
                 \langle\, \Phi_{0} \, |\,
                 {j^{t'\nu(dc)}_{\ell'
s'JM}}{}^{\dagger}(r')
                 \,|\, p(\Delta)h \, \rangle
                 \langle\, p(\Delta)h \, |\,
                 j^{t\nu(ab)}_{\ell sJM}(r)
                 \,|\, \Phi_{0} \, \rangle
                }
                {\omega +
\left({\epsilon}^{(a)}_{p(\Delta)}
                        - {\epsilon}^{}_{h}\right) - i\eta}
           \right],
           \nonumber \\
\label{eq:NC}
\end{eqnarray}
where
$\,|\, \Phi_{0} \, \rangle$ is the ground state of
$H^{(0)}$
and
$\,|\, p(\Delta)h \, \rangle$ denotes
a $ph (\Delta h)$ state.
Here
${\epsilon}^{}_{h}$
denotes
the single-hole energy
and
${\epsilon}^{(a)}_{p(\Delta)}$
is
the single-particle energy of the type $(a)$ particle.

In a continuum RPA
\cite{Shlomo75},
the sum over the discrete and continuum particle states
is carried out by use of a single-particle Green's
function,
\begin{eqnarray}
g^{(a)}_{}(\mbox{\boldmath $r$},\mbox{\boldmath
$r$}';\epsilon)
   &=&  \langle\, \mbox{\boldmath $r$} \, |\,
\frac{1}{\epsilon - h^{(a)} + i \eta} \,|\, \mbox{\boldmath
$r$}' \, \rangle
   \\
      &=&  \sum_{\ell jmm^{t}}
           {\cal Y}^{}_{\ell sjm}(\Omega_{r})
           \eta^{t}_{m^{t}}
           \frac{g^{(a)}_{\ell sjtm^{t}}(r,r';\epsilon)}
                {rr'}
           \left[
           {\cal Y}^{}_{\ell sjm}(\Omega_{r'})
           \eta^{t}_{m^{t}}
           \right]^{\dagger}
           ,
\end{eqnarray}
where $h^{(a)}$ is the single-particle Hamiltonian of
the type $(a)$ particle
and
${\cal Y}^{}_{\ell sjm}(\Omega_{r})
= \left[Y^{\ell}_{}(\Omega_{r})
  \otimes \chi^{s}_{}\right]^{j}_{m}$
with
the spin function
$\chi^{s}_{m^{s}}$
and
the isospin function
$\eta^{t}_{m^{t}}$.

Using
$g^{(a)}_{}(\mbox{\boldmath $r$},\mbox{\boldmath
$r$}';\epsilon)$
and
the single-hole wave function $u^{}_{n\beta}(r)$,
we get
\cite{Alberico85,Ichimura89}
\begin{eqnarray}
\lefteqn{
\hskip -1em
{\Pi^{(0)}}^{tt'\nu(ab,cd)}_{J\ell s \ell' s'}(r,r';\omega)
   =
       \delta_{ad} \delta_{bc}
       \sum_{\alpha}
       \sum_{(n\beta)\in occ}
} \nonumber \\
       && \hskip 9em
       \times
       \left[
       \delta_{aN}
       {\cal{B}}^{t\nu(ab)}_{\ell sJ}
       (\beta,\alpha)
       \frac{
             u^{}_{n\beta}{}^{\ast}(r)
            }
            {r^{2}}
       g^{(b)}_{\alpha}(r,r';\omega +
{\epsilon}^{}_{n\beta})
       \frac{
             u^{}_{n\beta}(r')
            }
            {r'^{2}}
       {\cal{B}}^{t'\nu(ab)}_{\ell' s'J'}{}^{\ast}
       (\beta,\alpha)
       \right. \nonumber \\
       && \hskip 9.3em
       \left.
       +
       \delta_{bN}
       {\cal{B}}^{t\nu(ab)}_{\ell stJ}
       (\alpha,\beta)
       \frac{
             u^{}_{n\beta}{}^{\ast}(r')
            }
            {r'^{2}}
       g^{(a)}_{\alpha}(r',r;-\omega +
{\epsilon}^{}_{n\beta})
       \frac{
             u^{}_{n\beta}(r)
            }
            {r^{2}}
       {\cal{B}}^{t'\nu(ab)}_{\ell' s'J'}{}^{\ast}
       (\alpha,\beta)
       \right]
       ,
       \nonumber \\
\end{eqnarray}
where
$\alpha = (\ell_{\alpha} s_{\alpha} j_{\alpha}
           t_{\alpha} m_{\alpha}^{t})$ and
$\beta  = (\ell_{\beta} s_{\beta} j_{\beta}
           t_{\beta} m_{\beta}^{t})$,
and
\begin{eqnarray}
{\cal{B}}^{t\nu(ab)}_{\ell sJ}
(\alpha_{1},\alpha_{2})
   &\equiv&
   \langle\, t_{1}m^{t}_{1} \, |\, {\tau}^{t(ab)}_{\nu}
\,|\, t_{2}m^{t}_{2} \, \rangle
   \sqrt{2j_{1}+1} \sqrt{2j_{2}+1}
   \nonumber \\
   &&
   \times
   \left\{\begin{array}{ccc} \ell_{1} & s_{1} & j_{1} \\
\ell & s & J \\ \ell_{2} & s_{2} & j_{2}
\end{array}\right\}
   \langle\, \ell_{1} \, \|\, i^{\ell} Y^{\ell}_{} \,\|\,
\ell_{2} \, \rangle
   \langle\, s_{1} \, \|\, {\sigma}^{s(ab)}_{} \,\|\, s_{2}
\, \rangle
   ,
\end{eqnarray}
with
\begin{eqnarray}
\langle\, s_{1} \, \|\, {\sigma}^{s(ab)}_{} \,\|\, s_{2} \,
\rangle
   &\equiv&
   \left\{ \begin{array}{ll}
   \delta_{s_{1} s_{2}} \sqrt{2s_{1}+1}  & \hskip 4em {\rm
for} \hskip 1em s=0 \\
   \sqrt{6}        & \hskip 4em {\rm for } \hskip 1em
s=1,\; s_{1} = s_{2} = 1/2 \\
   2               & \hskip 4em {\rm for } \hskip 1em
s=1,\; s_{1} = 3/2,\; s_{2} = 1/2 \\
   -2              & \hskip 4em {\rm for } \hskip 1em
s=1,\; s_{1} = 1/2,\; s_{2} = 3/2
   \end{array}
   \right.
   .
\end{eqnarray}

\subsection{Ring approximation}
We take into account the nuclear correlation
by the ring approximation.
We assume that the effective interaction is
the charge-independent local two-body force,
which is written as
\begin{eqnarray}
V^{(ab,cd)}_{}(\mbox{\boldmath $r$}-\mbox{\boldmath
$r$}';\omega)
      &=&
           \sum_{ss' \ell \ell'}
           \sum_{JM t\nu}
           \int^{\infty}_{0} {r_{1}}^{2}dr_{1}
{r_{2}}^{2}dr_{2} \;
           {j^{t\nu(ba)}_{\ell
sJM}}{}^{\dagger}(r_{1};\mbox{\boldmath $r$})
           W^{t(ab,cd)}_{J\ell s \ell'
s'}(r_{1},r_{2};\omega)
           j^{t\nu(cd)}_{\ell' s'JM}(r_{2};\mbox{\boldmath
$r$}')
        \label{eq:AC}
           .
           \nonumber \\
\end{eqnarray}
Then the polarization propagator satisfies the ring equation
\cite{Fetter71},
\begin{eqnarray}
\Pi^{tt'\nu}_{J\ell s \ell' s'}(r,r';\omega)
   &=&
        {\Pi^{(0)}}^{tt'\nu}_{J\ell s \ell' s'}(r,r';\omega)
   \nonumber \\
   &&
   +
   \sum_{t_{1} \ell_{1} s_{1} \ell_{2} s_{2}}
   \int^{\infty}_{0} {r_{1}}^{2}dr_{1}\, {r_{2}}^{2}dr_{2}
   \;
   {\Pi^{(0)}}^{tt_{1}\nu}_{J\ell s \ell_{1}
s_{1}}(r,r_{1};\omega)
   W^{t_{1}}_{J\ell_{1} s_{1} \ell_{2}
s_{2}}(r_{1},r_{2};\omega)
   \nonumber \\
   && \hskip 20.5em
   \times
   \Pi^{t_{1}t'\nu}_{J\ell_{2} s_{2} \ell'
s'}(r_{2},r';\omega)
   \label{eq:ring}
   ,
\end{eqnarray}
where $\Pi^{}_{}$ and $W^{}_{}$
are the matrices with respect to the particle types,
\begin{eqnarray}
   \Pi^{}_{}
   &\equiv&
   \left[
   \begin{array}{ccc}
     \Pi^{(N      N      ,N      N     )}_{}
     &
     \Pi^{(N      N      ,N      \Delta)}_{}
     &
     \Pi^{(N      N      ,\Delta N     )}_{}
     \\
     \Pi^{(N      \Delta ,N      N     )}_{}
     &
     \Pi^{(N      \Delta ,N      \Delta)}_{}
     &
     \Pi^{(N      \Delta ,\Delta N     )}_{}
     \\
     \Pi^{(\Delta N      ,N      N     )}_{}
     &
     \Pi^{(\Delta N      ,N      \Delta)}_{}
     &
     \Pi^{(\Delta N      ,\Delta N     )}_{}
   \end{array}
   \right]
   ,\\
   W^{}_{}
   &\equiv&
   \left[
   \begin{array}{cccc}
     W^{(N      N      ,N      N     )}_{}
     &
     W^{(N      N      ,N      \Delta)}_{}
     &
     W^{(N      N      ,\Delta N     )}_{}
     \\
     W^{(N      \Delta ,N      N     )}_{}
     &
     W^{(N      \Delta ,N      \Delta)}_{}
     &
     W^{(N      \Delta ,\Delta N     )}_{}
     \\
     W^{(\Delta N      ,N      N     )}_{}
     &
     W^{(\Delta N      ,N      \Delta)}_{}
     &
     W^{(\Delta N      ,\Delta N     )}_{}
   \end{array}
   \right]
   .
\end{eqnarray}

Using the symmetry of $W^{}_{}$,
we
introduce the grouped notation
$W^{[\alpha \beta]}_{}$
for
$W^{(ab,cd)}_{}$
as
\begin{eqnarray}
W^{[NN]}_{}            &\equiv& W^{(NN,NN)}_{},
\label{eq:WNN} \\
W^{[N\Delta]}_{}       &=&      W^{[\Delta N]}_{}
                                  \equiv
                                  W^{(NN,N\Delta)}_{}
                                  =
                                  W^{(NN,\Delta N)}_{}
                                  =
                                  W^{(N\Delta, NN)}_{}
                                  =
                                  W^{(\Delta N,NN)}_{},
\label{eq:WND} \\
W^{[\Delta\Delta]}_{}  &\equiv& W^{(N \Delta,N \Delta)}_{}
                                  =
                                  W^{(N \Delta,\Delta N)}_{}
                                  =
                                  W^{(\Delta N,N \Delta)}_{}
                                  =
                                  W^{(\Delta N,\Delta N)}_{}
                                  , \label{eq:WDD}
\end{eqnarray}
and correspondingly
the grouped representation of $\Pi^{}_{}$
as
\begin{eqnarray}
\Pi^{[NN]}_{}            &\equiv& \Pi^{(NN,NN)}_{}, \\
\Pi^{[N\Delta]}_{}       &\equiv& \Pi^{(NN,N\Delta)}_{}
                                   +
                                   \Pi^{(NN,\Delta N)}_{},
\\
\Pi^{[\Delta N]}_{}      &\equiv& \Pi^{(N\Delta, NN)}_{}
                                   +
                                   \Pi^{(\Delta N, NN)}_{},
\\
\Pi^{[\Delta\Delta]}_{}  &\equiv& \Pi^{(N \Delta,N
\Delta)}_{}
                                   +
                                   \Pi^{(N \Delta,\Delta
N)}_{}
                                   +
                                   \Pi^{(\Delta N,N
\Delta)}_{}
                                   +
                                   \Pi^{(\Delta N,\Delta
N)}_{}.
\end{eqnarray}
Then we can reduce the dimension of
$\Pi^{}_{}$ and $W^{}_{}$ in Eq.~(\ref{eq:ring}),
and express them as
\cite{Shiino86},
\begin{eqnarray}
   \Pi^{}_{}
   \equiv
   \left[
   \begin{array}{cc}
     \Pi^{[N      N     ]}_{}
     &
     \Pi^{[N      \Delta]}_{}
     \\
     \Pi^{[\Delta N     ]}_{}
     &
     \Pi^{[\Delta \Delta]}_{}
   \end{array}
   \right]
   ,
   \hskip 4em
   W^{}_{}
   \equiv
   \left[
   \begin{array}{cc}
     W^{[N      N     ]}_{}
     &
     W^{[N      \Delta]}_{}
     \\
     W^{[\Delta N     ]}_{}
     &
     W^{[\Delta \Delta]}_{}
   \end{array}
   \right]
   .
\end{eqnarray}
The ring equation
(\ref{eq:ring})
is now symbolically written as
\begin{eqnarray}
   \Pi^{}_{}
   =
   {\Pi^{(0)}}^{}_{} + {\Pi^{(0)}}^{}_{} W^{}_{} \Pi^{}_{}
   ,
\label{eq:NB}
\end{eqnarray}
with
\begin{eqnarray}
   {\Pi^{(0)}}^{}_{}
   =
   \left[
   \begin{array}{cc}
     {\Pi^{(0)}}^{[N      N     ]}_{}
     &
     0
     \\
     0
     &
     {\Pi^{(0)}}^{[\Delta \Delta]}_{}
   \end{array}
   \right]
   .
\end{eqnarray}
We also get the symmetry,
$\Pi^{[N\Delta]}_{} = \Pi^{[\Delta N]}_{}$,
since $W^{[N\Delta]}_{} = W^{[\Delta N]}_{}$.

\subsection{Spin-longitudinal and -transverse modes}
In this paper
we restrict ourselves to
the response functions for
the isovector spin-dependent currents ($s=t=1$).
So we use the abbreviations,
\begin{eqnarray}
j^{\nu(ab)}_{\ell JM}
\equiv
   j^{t=1 \nu(ab)}_{\ell s=1 JM}
   ,\;\;\;
\Pi^{\nu(ab,cd)}_{J\ell\ell'}
\equiv
   \Pi^{t=1 t'=1 \nu(ab,cd)}_{J\ell s=1 \ell' s'=1}
   .
\end{eqnarray}
We define
the spin-longitudinal
and
-transverse currents,
${j_{L}}^{}_{}$
and
${j_{T}}^{}_{}$,
in the momentum representation
as
\begin{eqnarray}
   {j_{L}}^{\nu(ab)}_{}(\mbox{\boldmath $q$})
   &\equiv&
       \sum^{A}_{k=1}
       {j_{L}}^{\nu(ab)}_{}(\mbox{\boldmath
$q$};\mbox{\boldmath $r$}_{k})
       \label{eq:AP}
       , \\
   {j_{T}}^{\nu(ab)}_{\mu}(\mbox{\boldmath $q$})
   &\equiv&
       \sum^{A}_{k=1}
       {j_{T}}^{\nu(ab)}_{\mu}(\mbox{\boldmath
$q$};\mbox{\boldmath $r$}_{k})
       \label{eq:AQ}
       ,
\end{eqnarray}
where
\begin{eqnarray}
   {j_{L}}^{\nu(ab)}_{}(\mbox{\boldmath
$q$};\mbox{\boldmath $r$}_{k})
      &\equiv&
      \left[\left(\tau^{t=1(ab)}_{\nu}\right)_{k}
      \left({\mbox{\boldmath $\sigma$}_{k}}^{(ab)}_{} \cdot
\mbox{$ \widehat{\mbox{\boldmath $q$}}$}\right)\right]
      {\rm e}^{-i\mbox{\boldmath $q$} \cdot \mbox{\boldmath
$r$}_{k}}
      , \\
   {j_{T}}^{\nu(ab)}_{\mu}(\mbox{\boldmath
$q$};\mbox{\boldmath $r$}_{k})
      &\equiv&
      \left[\left(\tau^{t=1(ab)}_{\nu}\right)_{k}
      \left({\mbox{\boldmath $\sigma$}_{k}}^{(ab)}_{}
\times \mbox{$ \widehat{\mbox{\boldmath
$q$}}$}\right)_{\mu}\right]
      {\rm e}^{-i\mbox{\boldmath $q$} \cdot \mbox{\boldmath
$r$}_{k}}
      .\label{eq:YD}
\end{eqnarray}
They are separated into the angle and radial parts of
$\mbox{\boldmath $q$}$ as
\cite{Alberico85}
\begin{eqnarray}
   {j_{L}}^{\nu(ab)}_{}(\mbox{\boldmath
$q$};\mbox{\boldmath $r$}_{k})
      &=&       \sum_{\ell JM}
                F^{L}_{\ell JM}(\Omega_{q})
                j^{\nu(ab)}_{\ell JM}(q;\mbox{\boldmath
$r$}_{k})
      \label{eq:AAL}
      ,\\
   {j_{T}}^{\nu(ab)}_{\mu}(\mbox{\boldmath
$q$};\mbox{\boldmath $r$}_{k})
      &=&       \sum_{\ell JM}
                F^{T}_{\ell JM\mu}(\Omega_{q})
                j^{\nu(ab)}_{\ell JM}(q;\mbox{\boldmath
$r$}_{k})
      \label{eq:AAT}
      ,
\end{eqnarray}
where
\begin{eqnarray}
   F^{L}_{\ell JM}(\Omega_{q})
      &\equiv&
      \left[4\pi a_{J\ell}
Y^{J}_{M}(\Omega_{q})\right]^{\ast}
      \label{eq:AH}
    , \\
   F^{T}_{\ell JM\mu}(\Omega_{q})
      &\equiv&
      \sum_{KQ}
      \left[4\pi i \cdot b_{JK\ell} (-)^{M+\ell}
            \left(\begin{array}{ccc} J & K & 1 \\ M & -Q &
-\mu \end{array}\right)
            Y^{K}_{Q}(\Omega_{q})\right]^{\ast}
      \label{eq:AI}
      ,
\end{eqnarray}
with
\begin{eqnarray}
   a_{J\ell}
   &\equiv&
       \langle\,J010\,|\,\ell 0\,\rangle
       ,\\
   b_{JK\ell}
   &\equiv&
       \sqrt{6 (2J+1) (2K+1)}
       \; a_{K\ell}
       \left\{\begin{array}{ccc} 1 & 1 & 1 \\ J & K & \ell
\end{array}\right\}
       .
\end{eqnarray}
The current
$j^{}_{}(q;\mbox{\boldmath $r$}_{k})$
is related with
$j^{}_{}(r;\mbox{\boldmath $r$}_{k})$
defined in Eq.~(\ref{eq:ZA})
as
\begin{eqnarray}
j^{\nu(ab)}_{\ell s JM}(q;\mbox{\boldmath $r$}_{k})
   &=&
   \int^{\infty}_{0}r^{2}dr\;
   j^{\nu (ab)}_{\ell s JM}(r;\mbox{\boldmath $r$}_{k})
j_{\ell}(qr)
   \label{eq:AB}
   ,
\end{eqnarray}
with the $\ell$-th order spherical Bessel function
$j_{\ell}(qr)$.

In terms of these currents,
we define
the spin-longitudinal
and
-transverse polarization propagators as
\begin{eqnarray}
   {\Pi_{L}}^{\nu (ab,cd)}_{} (\mbox{\boldmath $q$},
\mbox{\boldmath $q$}';\omega)
     &=&   \langle\, \Psi_{0} \, |\,
          \left[{\mbox{$ \widetilde{j}$}_{L}}^{\nu (ab)}_{}
(\mbox{\boldmath $q$})
           \frac{1}{\omega - (H - E_{0}) + i\eta}
           {{\mbox{$ \widetilde{j}$}_{L}}^{\nu
(dc)}_{}}{}^{\dagger} (\mbox{\boldmath $q$}')
       \right. \nonumber \\ && \hskip 2em \left.
         - {{\mbox{$ \widetilde{j}$}_{L}}^{\nu
(dc)}_{}}{}^{\dagger} (\mbox{\boldmath $q$}')
           \frac{1}{\omega + (H - E_{0}) - i\eta}
           {\mbox{$ \widetilde{j}$}_{L}}^{\nu (ab)}_{}
(\mbox{\boldmath $q$})
          \right]
           \,|\, \Psi_{0} \, \rangle
       \label{eq:AF}
           , \\
   {\Pi_{T}}^{\nu (ab,cd)}_{\mu\mu'} (\mbox{\boldmath $q$},
\mbox{\boldmath $q$}';\omega)
     &=&   \langle\, \Psi_{0} \, |\,
         \left[{\mbox{$ \widetilde{j}$}_{T}}^{\nu
(ab)}_{\mu} (\mbox{\boldmath $q$})
           \frac{1}{\omega - (H - E_{0}) + i\eta}
           {{\mbox{$ \widetilde{j}$}_{T}}^{\nu
(dc)}_{\mu'}}{}^{\dagger} (\mbox{\boldmath $q$}')
       \right. \nonumber \\ && \hskip 2em \left.
         - {{\mbox{$ \widetilde{j}$}_{T}}^{\nu
(dc)}_{\mu'}}{}^{\dagger} (\mbox{\boldmath $q$}')
           \frac{1}{\omega + (H - E_{0}) - i\eta}
           {\mbox{$ \widetilde{j}$}_{T}}^{\nu (ab)}_{\mu}
(\mbox{\boldmath $q$})
          \right]
           \,|\, \Psi_{0} \, \rangle
       \label{eq:AG}
           .
\end{eqnarray}
{}From Eqs.~(\ref{eq:AAL}), (\ref{eq:AAT}) and (\ref{eq:AB}),
they can be rewritten as
\begin{eqnarray}
   {\Pi_{L}}^{\nu(ab,cd)}_{}(\mbox{\boldmath $q$},
\mbox{\boldmath $q$}';\omega)
     &=& \sum_{JM \ell\ell'}
         F^{L}_{\ell JM}(\Omega_{q})
         F^{L}_{\ell' JM}{}^{\ast}(\Omega_{q'})
         \Pi^{\nu(ab,cd)}_{J\ell\ell'}(q,q';\omega)
         , \label{eq:QA}
         \\
   {\Pi_{T}}^{\nu(ab,cd)}_{\mu\mu'}(\mbox{\boldmath $q$},
\mbox{\boldmath $q$}';\omega)
     &=& \sum_{JM \ell\ell'}
         F^{T}_{\ell JM\mu}(\Omega_{q})
         F^{T}_{\ell' JM\mu'}{}^{\ast}(\Omega_{q'})
         \Pi^{\nu(ab,cd)}_{J\ell\ell'}(q,q';\omega)
         ,\label{eq:QB}
\end{eqnarray}
where
$\Pi^{}_{}(q,q';\omega)$ is related with
$\Pi^{}_{}(r,r';\omega)$ defined in Eq.~(\ref{eq:ZB}) as

\begin{eqnarray}
   \Pi^{\nu(ab,cd)}_{J\ell \ell'}(q,q';\omega)
     &=&   \int^{\infty}_{0} \! r^{2}dr\; r'^{2}dr' \;
           j_{\ell}(qr)
           \Pi^{\nu(ab,cd)}_{J\ell \ell'}(r,r';\omega)
           j_{\ell'}(q'r')
           .
\end{eqnarray}

\subsection{($\pi + \rho + g'_{}$) model}
For the effective interaction (\ref{eq:AC}),
the ($\pi + \rho + g'_{}$) model
is commonly adopted.
It gives
\begin{eqnarray}
   V^{(ab,cd)}_{}(\mbox{\boldmath $r$}_{1}-\mbox{\boldmath
$r$}_{2};\omega)
   =
   {V_{L}}^{(ab,cd)}_{}(\mbox{\boldmath
$r$}_{1}-\mbox{\boldmath $r$}_{2};\omega)
   +
   {V_{T}}^{(ab,cd)}_{}(\mbox{\boldmath
$r$}_{1}-\mbox{\boldmath $r$}_{2};\omega)
   ,\label{eq:MA}
\end{eqnarray}
with
\begin{eqnarray}
   {V_{L}}^{(ab,cd)}_{}(\mbox{\boldmath
$r$}_{1}-\mbox{\boldmath $r$}_{2};\omega)
      &\equiv&
      \int \frac{d^{3}q}{(2\pi)^{3}}\;
      {\rm e}^{i\mbox{\boldmath $q$} \cdot (\mbox{\boldmath
$r$}_{1}-\mbox{\boldmath $r$}_{2})}
      {W_{L}}^{(ab,cd)}_{}(q,\omega) \nonumber \\
      &&\hskip 4em
      \times
      \left[{\mbox{\boldmath $\tau$}_{1}}^{(ab)}_{} \cdot
{\mbox{\boldmath $\tau$}_{2}}^{(cd)}_{}\right]
      \left[{\mbox{\boldmath $\sigma$}_{1}}^{(ab)}_{} \cdot
\mbox{$ \widehat{\mbox{\boldmath $q$}}$}\right]
      \left[{\mbox{\boldmath $\sigma$}_{2}}^{(cd)}_{} \cdot
\mbox{$ \widehat{\mbox{\boldmath $q$}}$}\right], \\
   {V_{T}}^{(ab,cd)}_{}(\mbox{\boldmath
$r$}_{1}-\mbox{\boldmath $r$}_{2};\omega)
      &\equiv& \int \frac{d^{3}q}{(2\pi)^{3}}\;
      {\rm e}^{i\mbox{\boldmath $q$} \cdot (\mbox{\boldmath
$r$}_{1}-\mbox{\boldmath $r$}_{2})}
      {W_{T}}^{(ab,cd)}_{}(q,\omega) \nonumber \\
      &&\hskip 4em
      \times
      \left[{\mbox{\boldmath $\tau$}_{1}}^{(ab)}_{} \cdot
{\mbox{\boldmath $\tau$}_{2}}^{(cd)}_{}\right]
      \left[\left({\mbox{\boldmath $\sigma$}_{1}}^{(ab)}_{}
\times \mbox{$ \widehat{\mbox{\boldmath $q$}}$}\right) \cdot
      \left({\mbox{\boldmath $\sigma$}_{2}}^{(cd)}_{}
\times \mbox{$ \widehat{\mbox{\boldmath
$q$}}$}\right)\right].
\end{eqnarray}

In the grouped representation
(\ref{eq:WNN})-(\ref{eq:WDD}),
${W_{L}}^{}_{}$
and
${W_{T}}^{}_{}$
are given
\cite{Alberico82,Oset82,Shiino86}
as
\begin{eqnarray}
   {W_{L}}^{[\alpha\beta]}_{}(q,\omega)
   &=&
       \frac{f_{\alpha}f_{\beta}}{m_{\pi}^{2}}
       \left[
       g'_{\alpha\beta}(q)
       +
       \Gamma^{\pi}_{\alpha}(q,\omega)
       \Gamma^{\pi}_{\beta}(q,\omega)
       \frac{q^{2}}{\omega^{2} - q^{2} - m_{\pi}^{2}}
       \right],
       \\
   {W_{T}}^{[\alpha\beta]}_{}(q,\omega)
   &=&
       \frac{f_{\alpha}f_{\beta}}{m_{\pi}^{2}}
       \left[
       g'_{\alpha\beta}(q)
       +
       C^{\rho}_{\alpha\beta}
       \Gamma^{\rho}_{\alpha}(q,\omega)
       \Gamma^{\rho}_{\beta}(q,\omega)
       \frac{q^{2}}{\omega^{2} - q^{2} - m_{\rho}^{2}}
       \right],
\end{eqnarray}
where
$\alpha$ and $\beta$ denote $N$ or $\Delta$,
and
$m_{\pi}$
and
$m_{\rho}$
are the pion and the $\rho$-meson masses, respectively.
The coefficient
$
C^{\rho}_{\alpha\beta} \equiv
\frac{f^{\rho}_{\alpha}f^{\rho}_{\beta}}{m_{\rho}^{2}}
/
\frac{f_{\alpha}f_{\beta}}{m_{\pi}^{2}}
$
is the ratio of the $\rho$-meson exchange coupling
to the $\pi$-meson one.
The following vertex form factors are used,
\begin{eqnarray}
   \Gamma^{\pi}_{\alpha}(q,\omega)
   =
       \frac{
             m_{\pi}^{2}
             -
             \Lambda_{\pi}^{2}
            }
            {
             \omega^{2}
             -
             q^{2}
             -
             \Lambda_{\pi}^{2}
            }
       ,\;\;\;\;
   \Gamma^{\rho}_{\alpha}(q,\omega)
   =
       \frac{
             m_\rho^{2}
             -
             \Lambda_{\rho}^{2}
            }
            {
             \omega^{2}
             -
             q^{2}
             -
             \Lambda_{\rho}^{2}
            }
       .
\end{eqnarray}
The Landau-Migdal parameters
$g'_{}(q)$'s
depend on $q$ so weakly that we neglect their $q$
dependence
in the region of $q \leq 3 \;{\rm fm}^{-1}$
\cite{Dickhoff87,Brown93}.
They are treated as free parameters.
The effective interaction (\ref{eq:MA}) is expressed
in angular-momentum representation as
\begin{eqnarray}
   V^{(ab,cd)}_{}(\mbox{\boldmath $r$}_{1}-\mbox{\boldmath
$r$}_{2};\omega)
      =    \sum_{JM\nu \ell \ell'}
           \frac{2}{\pi} \int^{\infty}_{0} q^{2}dq \;
           {j^{\nu(ba)}_{\ell
JM}}{}^{\dagger}(q;\mbox{\boldmath $r$}_{1})
           W^{(ab,cd)}_{J\ell \ell'}(q,\omega)
           j^{\nu(cd)}_{\ell' JM}(q;\mbox{\boldmath
$r$}_{2})
           ,
\end{eqnarray}
where
\begin{eqnarray}
   W^{(ab,cd)}_{J\ell \ell'}(q,\omega)
   =
       {W_{L}}^{(ab,cd)}_{J\ell \ell'}(q,\omega)
       +
       {W_{T}}^{(ab,cd)}_{J\ell \ell'}(q,\omega)
       ,
\end{eqnarray}
with
\begin{eqnarray}
{W_{L}}^{(ab,cd)}_{J\ell \ell'}(q,\omega)
  &\equiv&
  a_{J\ell} {W_{L}}^{(ab,cd)}_{}(q,\omega) a_{J\ell'}
  , \\
{W_{T}}^{(ab,cd)}_{J\ell \ell'}(q,\omega)
  &\equiv&
  {W_{T}}^{(ab,cd)}_{}(q,\omega) [\delta_{\ell\ell'} -
a_{J\ell}a_{J\ell'}]
  .
\end{eqnarray}
The coordinate representation of the effective interaction
in Eq.~(\ref{eq:AC}) is
given by
\begin{eqnarray}
   W^{(ab,cd)}_{J\ell \ell'}(r_{1},r_{2};\omega)
    \equiv
       \frac{2}{\pi} \int^{\infty}_{0} q^{2}dq \;
       j_{\ell}(qr_{1})
       W^{(ab,cd)}_{J\ell \ell'}(q,\omega)
       j_{\ell'}(qr_{2})
       .
\end{eqnarray}
The ring equation (\ref{eq:ring}) reads as
\begin{eqnarray}
\Pi^{\nu}_{J\ell\ell'}(r,r';\omega)
   &=&
        {\Pi^{(0)}}^{\nu}_{J\ell\ell'}(r,r';\omega)
   \nonumber \\
   &&   +
        \sum_{\ell_{1}\ell_{2}}
        \int^{\infty}_{0} {r_{1}}^{2}dr_{1}\,
{r_{2}}^{2}dr_{2}
        \;
        {\Pi^{(0)}}^{\nu}_{J\ell\ell_{1}}(r,r_{1};\omega)
        W^{}_{J\ell_{1}\ell_{2}}(r_{1},r_{2};\omega)
        \Pi^{\nu}_{J\ell_{2}\ell'}(r_{2},r';\omega)
        .
\end{eqnarray}

\subsection{Spin-longitudinal and -transverse response
functions}
We define
the spin-longitudinal and -transverse response functions as
\begin{eqnarray}
R^{\nu (ab,cd)}_{L}(q, \omega)
   &\equiv&
   \frac{1}{A}
   \sum_{n \neq 0}
   \langle\, \Psi_{0} \, |\, {j_{L}}^{\nu
(ab)}_{}(\mbox{\boldmath $q$}, \omega) \,|\, \Psi_{n} \,
\rangle
   \langle\, \Psi_{n} \, |\, {{j_{L}}^{\nu
(dc)}_{}}{}^{\dagger}(\mbox{\boldmath $q$}, \omega) \,|\,
\Psi_{0} \, \rangle
   \nonumber \\
   &&
   \times
   \delta[\omega - (E_{n} - E_{0})]
   ,\\
R^{\nu (ab,cd)}_{T}(q, \omega)
   &\equiv&
   \frac{1}{A}
   \sum_{n \neq 0}
   \frac{1}{2}
   \sum_{\mu}
   \langle\, \Psi_{0} \, |\, {j_{T}}^{\nu
(ab)}_{\mu}(\mbox{\boldmath $q$}, \omega) \,|\, \Psi_{n} \,
\rangle
   \langle\, \Psi_{n} \, |\, {{j_{T}}^{\nu
(dc)}_{\mu}}{}^{\dagger}(\mbox{\boldmath $q$}, \omega)
\,|\, \Psi_{0} \, \rangle
   \nonumber \\
   &&
   \times
   \delta[\omega - (E_{n} - E_{0})]
   .\label{eq:YC}
\end{eqnarray}
These are rewritten in the grouped representation as
\begin{eqnarray}
R^{\nu [\alpha \beta]}_{L}(q, \omega)
  &=&
   -
   \frac{1}{A}
   \frac{1}{\pi}
   \mbox{Im}
   {\Pi_{L}}^{\nu [\alpha \beta]}_{}(q, \omega)
   , \\
R^{\nu [\alpha \beta]}_{T}(q, \omega)
  &=&
   -
   \frac{1}{A}
   \frac{1}{\pi}
   \mbox{Im}
   {\Pi_{T}}^{\nu [\alpha \beta]}_{}(q, \omega)
   ,
\end{eqnarray}
with the momentum diagonal parts of
the polarization propagators (\ref{eq:AF}) and
(\ref{eq:AG}),
\begin{eqnarray}
   {\Pi_{L}}^{\nu[\alpha\beta]}_{} (q,\omega)
   &\equiv&
       {\Pi_{L}}^{\nu[\alpha\beta]}_{} (\mbox{\boldmath
$q$}, \mbox{\boldmath $q$};\omega)
       \nonumber \\
   &=&
       4\pi
       \sum_{J\ell\ell'}
       (2J+1)
       a_{J\ell} a_{J\ell'}
       \Pi^{\nu[\alpha\beta]}_{J\ell\ell'} (q,q;\omega)
       , \\
   {\Pi_{T}}^{\nu[\alpha\beta]}_{} (q;\omega)
   &\equiv&
       \frac{1}{2}
           \sum_{\mu}
           {\Pi_{T}}^{\nu[\alpha\beta]}_{\mu\mu}
(\mbox{\boldmath $q$}, \mbox{\boldmath $q$};\omega)
           \nonumber \\
   &=&
       2\pi
       \sum_{J\ell\ell'}
       (2J+1)
       (\delta_{\ell\ell'} - a_{J\ell} a_{J\ell'})
       \Pi^{\nu[\alpha\beta]}_{J\ell\ell'} (q,q;\omega)
        ,
\end{eqnarray}
where Eqs.~(\ref{eq:QA}),
(\ref{eq:QB}),
(\ref{eq:AH}) and
(\ref{eq:AI}) are used.

As to the responses for nucleon probes,
we assume that
the ratio of
the scattering amplitudes,
the $NN \rightarrow N\Delta$ to
the $NN \rightarrow NN$,
is $f_{\Delta}/f_{N}$,
as is commonly done
implicitly
\cite{Alberico82}
and explicitly
\cite{Shiino86}.
Then relevant response functions are given by
\begin{eqnarray}
R^{\nu}_{L,T}(q, \omega)
  &=&
   R^{\nu [NN]}_{L,T}(q, \omega)
   +
   2 \frac{f_{\Delta}}{f_{N}}
   R^{\nu [N\Delta]}_{L,T}(q, \omega)
   +
   \left(\frac{f_{\Delta}}{f_{N}}\right)^{2}
   R^{\nu [\Delta\Delta]}_{L,T}(q, \omega)
   \label{eq:RLRTtotal}
   ,
\end{eqnarray}
as was shown in Eq.~(\ref{eq:NA}).
The uncorrelated ones are
\begin{eqnarray}
R^{(0)\nu}_{L,T}(q, \omega)
  &=&
   R^{(0)\nu [NN]}_{L,T}(q, \omega)
   +
   \left(\frac{f_{\Delta}}{f_{N}}\right)^{2}
   R^{(0)\nu [\Delta\Delta]}_{L,T}(q, \omega)
   .
\end{eqnarray}
In the quasi-elastic region,
the second term does not contribute
since real $\Delta$ production does not occur.
For $(e,e')$ scattering, the cross section is
expressed
in the one-photon-exchange approximation
as
\begin{eqnarray}
\frac{d^{2}\sigma}{d\epsilon d\Omega}
   =
   \sigma_{M}
   \left[
   \left(\frac{q_{\mu}^{2}}{\mbox{\boldmath
$q$}^{2}}\right)^{2}
   S_{L}(q, \omega)
   +
   \left(\tan^{2}\frac{\theta}{2}
   -
   \frac{q_{\mu}^{2}}{2\mbox{\boldmath $q$}^{2}}\right)
   S_{T}(q, \omega)
   \right],
\end{eqnarray}
with
the Mott cross section $\sigma_{M}$,
a transferred four-momentum $q_{\mu}=(\omega,
\mbox{\boldmath $q$})$,
$q_{\mu}^{2} = \omega^{2} - \mbox{\boldmath $q$}^{2}$ and
a scattering angle $\theta$.
The dynamic structure factor
\cite{Alberico86},
$S_{L}$ and $S_{T}$,
are given by\footnote{
Some papers use the notation
$(4\pi/M_T) S_{L,T}$
instead of the present $S_{L,T}$,
where $M_{T}$ stands for the target mass.
}
\begin{eqnarray}
S_{L}(q, \omega)
   &=&
   \sum_{n \neq 0}
   \left|\langle\, \Psi_{n} \, |\,
{\rho_{C}}{}^{\dagger}(\mbox{\boldmath $q$}, \omega) \,|\,
\Psi_{0} \, \rangle\right|^{2}
   \delta[\omega - (E_{n} - E_{0})]
   ,\\
S_{T}(q, \omega)
   &=&
   \sum_{n \neq 0}
   \left|\langle\, \Psi_{n} \, |\, {\mbox{\boldmath
$J$}_{T}}{}^{\dagger}(\mbox{\boldmath $q$}, \omega) \,|\,
\Psi_{0} \, \rangle\right|^{2}
   \delta[\omega - (E_{n} - E_{0})]
   ,\label{eq:YB}
\end{eqnarray}
where $(\rho_{C}, \mbox{\boldmath $J$}_{T})$ is
the nuclear electromagnetic current operator.
In this paper,
we restrict ourselves to the transverse part
$\mbox{\boldmath $J$}_{T}$,
which is given by the sum of
the one-body convection
and
magnetic currents,
and
the exchange current,
$\mbox{\boldmath $J$}_{T} = \mbox{\boldmath $J$}^{\rm
conv}_{T} + \mbox{\boldmath $J$}^{\rm mag}_{T} +
\mbox{\boldmath $J$}^{\rm exch}_{T}$.
We neglect the convection current
since its contribution is small
\cite{Alberico84}.
Although the exchange current
contributes to a certain extent,
we do not take into account of this term
since we only consider the responses of one-body operators.
The magnetic current
is given by
\begin{eqnarray}
\mbox{\boldmath $J$}^{\rm mag}_{T}(\mbox{\boldmath $q$})
   &=&
   -
   \frac{i}{2m_{N}}
   \sum_{k}
   {\rm e}^{-i\mbox{\boldmath $q$} \cdot \mbox{\boldmath
$r$}_{k}}
   \left[
   \left\{
   G^{IS}_{\gamma NN}(q_{\mu}^{2})
   +
   G^{IV}_{\gamma NN}(q_{\mu}^{2})
   \left(\tau^{(NN)}_{0}\right)_{k}
   \right\}
   \left({\mbox{\boldmath $\sigma$}_{k}}^{(NN)}_{} \times
\mbox{\boldmath $q$}\right)
   \right.
   \nonumber \\
   &&
   +
   \left.
   G^{}_{\gamma N \Delta}(q_{\mu}^{2})
   \left\{
   \left(\tau^{(N \Delta)}_{0}\right)_{k}
   \left({\mbox{\boldmath $\sigma$}_{k}}^{(N \Delta)}_{}
\times \mbox{\boldmath $q$}\right)
   +
   \left(\tau^{(\Delta N)}_{0}\right)_{k}
   \left({\mbox{\boldmath $\sigma$}_{k}}^{(\Delta N)}_{}
\times \mbox{\boldmath $q$}\right)
   \right\}
   \right]
   ,
\end{eqnarray}
with the nucleon mass $m_{N}$.
We take the magnetic form factors in the form of
\begin{eqnarray}
G^{}_{\gamma\alpha\beta}(q_{\mu}^{2})
   \equiv
   f_{\gamma\alpha\beta}
   \left[
   1 - \frac{q_{\mu}^{2}}{\lambda^{2}}
   \right]^{-2}
   ,
\end{eqnarray}
where $\lambda = 855 {\;\rm MeV}/c$
\cite{Meziani85}
and
$f_{}$'s are the magnetic strengths.
Then we get
\begin{eqnarray}
\mbox{\boldmath $J$}^{\rm mag}_{T}(\mbox{\boldmath $q$})
   &=&
   -
   \frac{i}{2m_{N}}
   G^{IV}_{\gamma NN}(q_{\mu}^{2})
   \sum_{k}
   {\rm e}^{-i\mbox{\boldmath $q$} \cdot \mbox{\boldmath
$r$}_{k}}
   \left[
   \left\{
   \frac{f^{IS}_{\gamma NN}}{f^{IV}_{\gamma NN}}
   +
   \left(\tau^{(NN)}_{0}\right)_{k}
   \right\}
   \left({\mbox{\boldmath $\sigma$}_{k}}^{(NN)}_{} \times
\mbox{\boldmath $q$}\right)
   \right.
   \nonumber \\
   &&
   +
   \left.
   \frac{f_{\gamma N \Delta}}{f^{IV}_{\gamma NN}}
   \left\{
   \left(\tau^{(N \Delta)}_{0}\right)_{k}
   \left({\mbox{\boldmath $\sigma$}_{k}}^{(N \Delta)}_{}
\times \mbox{\boldmath $q$}\right)
   +
   \left(\tau^{(\Delta N)}_{0}\right)_{k}
   \left({\mbox{\boldmath $\sigma$}_{k}}^{(\Delta N)}_{}
\times \mbox{\boldmath $q$}\right)
   \right\}
   \right]
   \label{eq:YA}
   ,
\end{eqnarray}
where
\begin{eqnarray}
   f^{IS}_{\gamma NN} = \left(\mu_p + \mu_n\right)/2,
\hskip 2em
   f^{IV}_{\gamma NN} = \left(\mu_p - \mu_n\right)/2,
\end{eqnarray}
with
$\mu_{p} =  2.79$ and
$\mu_{n} = -1.91$.
When we neglect the isospin-mixing
and consider only $T=0$ target nuclei,
the isoscalar and isovector parts do not interfere.
Then we can neglect the isoscalar part
since
$\left(f^{IS}_{\gamma NN}/f^{IV}_{\gamma NN}\right)^{2}
\approx 0.04$
is very small.
Inserting Eq.~(\ref{eq:YA}) into Eq.~(\ref{eq:YB})
and using Eqs.~(\ref{eq:YC}) and (\ref{eq:YD}),
we get
\begin{eqnarray}
S_{T}(q, \omega)
   &=&
   \sum_{n \neq 0}
   \left|\langle\, \Psi_{n} \, |\, {\mbox{\boldmath
$J$}^{\rm mag}_{T}}{}^{\dagger}(\mbox{\boldmath $q$},
\omega) \,|\, \Psi_{0} \, \rangle\right|^{2}
   \delta[\omega - (E_{n} - E_{0})]
   \\
   &=&
   2A
   \left|
   \frac{q}{2m_{N}}
   G^{IV}_{\gamma NN}(q_{\mu}^{2})
   \right|^{2}
   R^{(e,e')}_{T}(q, \omega),
\end{eqnarray}
where
\begin{eqnarray}
R^{(e,e')}_{T}(q, \omega)
   \equiv
   R^{\nu=0 [NN]}_{T}(q, \omega)
   +
   2
   \frac{f_{\gamma N \Delta}}{f^{IV}_{\gamma NN}}
   R^{\nu=0 [N\Delta]}_{T}(q, \omega)
   +
   \left(
   \frac{f_{\gamma N \Delta}}{f^{IV}_{\gamma NN}}
   \right)^{2}
   R^{\nu=0 [\Delta\Delta]}_{T}(q, \omega)
   \label{eq:Ree}
   .
\end{eqnarray}

\section{Effects of the $\Delta$-Hole Mixing}
\label{sec:DeltaHole}

In the following two sections
we present our numerical calculations
of the isovector spin-longitudinal and -transverse
response functions,
$R^{}_{L}(q,\omega)$ and
$R^{}_{T}(q,\omega)$,
for the doubly (sub-)closed shell nuclei,
$^{40}$Ca, $^{16}$O and $^{12}$C,
and analyze them from various points of view.

In this section
we compare
$R^{}_{L,T}$ with and without $\Delta$
to see effects of $\Delta$-mixing,
and investigate relative importance of
the process-decomposed response functions
$R^{[\alpha\beta]}_{L,T}$.
This manifests the $\Delta$ effects more clearly.
We also discuss the energy-weighted and energy-non-weighted
sums.

Calculations are carried out by the ring approximation,
the Tamm-Dancoff approximation (TDA) and
without any residual interactions.
These results will be called
the RPA, TDA and uncorrelated response functions,
respectively.

\subsection{Technical comments and choice of parameters}
Before presenting the numerical calculations,
we make some technical comments and
summarize the values of the parameters.

In the previous section,
$H$ and $\Psi$ include the center-of-mass motion,
however, it is better to replace them by the intrinsic ones
to isolate the structure part.
Then the transferred energy $\omega$ should be replaced by
that to the intrinsic state,
$\omega_{\rm int} = \omega - \omega_{\rm recoil}
              = \omega - q^2 / (2Am_N)$.
Similarly the transferred momentum $q$ is replaced by
that to the relative motion between the active nucleon
and the remaining $(A-1)$-nucleon core,
$q_{\rm int} = [(A-1)/A] q$.
For simplicity we suppress the script ``int''
in Sects.~\ref{sec:DeltaHole} and \ref{sec:EffectivInt}.

We took the single-particle potential for $N$ and $\Delta$
as
\begin{eqnarray}
   U(r)
   &=&
       -\left( V + iW  \right)
       \frac{1}
            {1 + \exp\left( \frac{r-R}{a} \right)}
       -2
       \frac{1}
            {m_{\pi}^{2}}
       \frac{
            V_{\ell s}
            }
            {a}
       \frac{
             \exp\left( \frac{r-R}{a} \right)
            }
            {
             r\left[1 + \exp\left( \frac{r-R}{a} \right)
\right]^{2}
            }
       \left( \mbox{\boldmath $\ell$} \cdot \mbox{\boldmath
$s$} \right)
       +
       V_{\mbox{coul}},
\end{eqnarray}
with $R = r_{0} A^{1/3}$.
$V_{\mbox{coul}}$ is the Coulomb potential of
the uniformly charged sphere with the radius parameter
$r_{c}$.
The shape parameters are fixed to be
$r_{0} = r_{c} = 1.27 \;{\rm fm}$ and
$a = 0.67 \;{\rm fm}$
\cite{Bohr75}.
For the nucleon
the spin-orbit potential depth $V_{\ell s}$ are fixed to be
$6.5{\;\rm MeV}$  for $^{12}$C,
$10.4{\;\rm MeV}$   for $^{16}$O and
$10.0{\;\rm MeV}$ for $^{40}$Ca.
The real potential depth $V$ is so determined
as to give the observed
separation energy of the outermost occupied state.
The imaginary potential depth $W$ are fixed to be
zero for the occupied (hole) states and
$5.0 {\;\rm MeV}$ for the particle states.
For $\Delta$
we set
$V = 30 {\;\rm MeV}$ and
$W = V_{\ell s} = 0.0 {\;\rm MeV}$
since we do not have enough information
for such virtual $\Delta$ appearing in the quasi-elastic
region.

The masses are chosen to be
$m_{N} = 940 {\;\rm MeV}$,
$m_{\Delta} = 1236 {\;\rm MeV}$,
$m_{\pi} = 139 {\;\rm MeV}$, and
$m_{\rho} = 770 {\;\rm MeV}$
unless explicitly mentioned.
The coupling constants are fixed to be
$f_{N}{}^{2}/4\pi = 0.081$,
$f_{\Delta}/f_{N} = 2.00$, and
$C^{\rho}_{\alpha\beta} =2.18$.
The cutoff parameters are set to be
$\Lambda_{\pi}  = 1300 {\;\rm MeV}$ and
$\Lambda_{\rho} = 2000 {\;\rm MeV}$
\cite{Alberico82}.

\subsection{Effects of the $\Delta$ components}
\label{sec:DeltaEffect}

Here we present the energy spectra of the response
functions
at $q = 1.70 \;{\rm fm}^{-1}$
for the charge exchange mode related with
the $(p,n)$-like reactions ($\nu=-1$).
{}From now on we suppress $\nu$ on $R^{}_{}$.
The Landau-Migdal parameters are taken to be
$(g'_{NN}, g'_{N\Delta}, g'_{\Delta\Delta}) = (0.6, 0.4,
0.5)$.

In Fig.~\ref{fig:WFIG_A},
we show the RPA response functions
$R^{}_{L,T}$
of $^{40}$Ca
(a) without and (b) with $\Delta$.
The uncorrelated response functions
$R^{(0)}_{L,T}$
are also shown,
which are very close to each other for this $LS$ closed
shell.
Slight difference comes from the spin-orbit force.
Fluctuations seen at lower $\omega$ are somewhat artificial
because the spreading widths of the hole states are not
included
in the present calculation.

When $\Delta$ is not included,
$R^{}_{L}$ is enhanced below about 85MeV
but quenched above that.
On the other hand,
$R^{}_{T}$ is quenched below about 96MeV
and enhanced above that.
Consequently $R^{}_{L}$ is softened
but $R^{}_{T}$ is hardened.

Once $\Delta$ is introduced,
$R^{}_{L}$ increases but $R^{}_{T}$ decreases
for the whole quasi-elastic region
as seen in Fig.~\ref{fig:WFIG_A}(b).
This is attributed to the coupling interaction
$W^{[N\Delta]}$,
which brings down the spin-longitudinal strength
from the $\Delta h$ to the $ph$ sector
but brings up the spin-transverse strength
in the opposite direction.
This is the essential effect of $\Delta$.

Fig.~\ref{fig:WFIG_F} shows the response functions with
$\Delta$
for (a) $^{16}$O and (b) $^{12}$C.
Qualitative features are common for all the nuclei
and the RPA effects are stronger for larger $A$.
The difference
between
$R^{(0)}_{L}$ and
$R^{(0)}_{T}$
is larger for $^{12}$C than for other two nuclei,
because $^{12}$C is not the spin-saturated nucleus.

To see the situation in more detail,
we separate the response functions
into the process-decomposed ones.
The RPA response functions
$R^{}_{L}$ and $R^{}_{T}$ of $^{40}$Ca
are decomposed into
$R^{[\alpha\beta]}_{L}$ in Fig.~\ref{fig:WFIG_B}(a) and
$R^{[\alpha\beta]}_{T}$ in Fig.~\ref{fig:WFIG_B}(b).
The main contribution comes from $R^{[NN]}_{}$, but
that from $R^{[N\Delta]}_{}$ is also significant.
The contribution from $R^{[\Delta\Delta]}_{}$
is negligibly small.
An important point is that
$R^{[NN]}_{L}$ and $R^{[N\Delta]}_{L}$
contribute constructively,
whereas $R^{[NN]}_{T}$ and $R^{[N\Delta]}_{T}$
do destructively.
This explains
the shift of
the spin-longitudinal and -transverse strengths seen above.
To see the backward effect we show the TDA response
functions
without $\Delta$ in Fig.~\ref{fig:WFIG_N}.
Softening of $R^{}_{L}$ and hardening of $R^{}_{T}$
are well developed,
but their magnitudes do not change so much.
Comparing with Fig.~\ref{fig:WFIG_A}(a), we can say that
the backward amplitudes in the ring approximation induce
further enhancement of $R^{}_{L}$ and
quenching of $R^{}_{T}$.

These behaviors are qualitatively understood
in the following way.
The formal solution of the RPA equation
(\ref{eq:NB}) is given by
\begin{eqnarray}
\Pi^{}_{} =
\left[1-{\Pi^{(0)}}^{}_{}W\right]^{-1}{\Pi^{(0)}}^{}_{}, \
\ \ \
{\Pi^{(0)}}^{}_{} = \Pi^{(0)}_{FW} + \Pi^{(0)}_{BK}.
\label{eq:BB}
\end{eqnarray}
where $\Pi^{(0)}_{FW}$ and $\Pi^{(0)}_{BK}$
are the forward and the backward
part of the uncorrelated polarization propagator.
The fact to be kept in mind is that
around the present momentum $(q = 1.7\;{\rm fm}^{-1})$
the spin-longitudinal effective interaction
$W^{[\alpha\beta]}_{L}(q,\omega)$
is negative if $g'_{\alpha\beta} <0.7$
but
the spin-transverse one $W^{[\alpha\beta]}_{T}(q,\omega)$
is positive
if $g'_{\alpha\beta} > 0.25$ for the present parameters
(see Fig.~\ref{fig:WFIG_O}).

First let us consider the cases without $\Delta$.
For simplicity
we treat $\Pi^{}_{}$ and $W^{}_{}$ as c-numbers instead of
matrices
like in a Fermi gas model
and consider a real single-particle potential.
Then response functions are expressed as
\begin{eqnarray}
R^{}_{} = \left|1-{\Pi^{(0)}}^{}_{}W\right|^{-2}R^{(0)}_{}.
\end{eqnarray}
The real part of $\Pi^{(0)}_{FW}$
changes the sign in the middle of the energy range
concerned,
negative at lower side of $\omega$
but positive at higher side,
as is seen from the first term of r.h.s.\ of
Eq.~(\ref{eq:NC}).
Hence
$\Pi^{(0)}_{FW}W^{}_{T}$ is negative
but
$\Pi^{(0)}_{FW}W^{}_{L}$ positive
at lower $\omega$,
and thus
$R^{}_{T}$ is quenched
but
$R^{}_{L}$ enhanced
if $W^{}_{L}$ is not so strong
($\Pi^{(0)}_{FW}W^{}_{L} < 2$).
At higher $\omega$ the situation is opposite.
As the results the TDA response functions are
softened for the spin-longitudinal mode but
hardened for the spin-transverse mode.
On the other hand $\Pi^{(0)}_{BK}$ is always negative
(see Eq.~(\ref{eq:NC})).
Therefore $R^{}_{L}$ is enhanced but $R^{}_{T}$ quenched
by the backward amplitude in the whole energy region.

Next let us consider the cases with $\Delta$.
In the first order of $W^{}_{}$
the RPA response functions $R^{[N\Delta]}_{L,T}$ are given
by
\begin{eqnarray}
R^{[N\Delta]}_{L,T}
=
{\Pi^{(0)}}^{[\Delta \Delta]}_{}
{W_{L,T}}^{[N\Delta]}_{}
R^{(0)[NN]}_{L,T}.
\end{eqnarray}
The uncorrelated $\Delta h$ polarization propagator
${\Pi^{(0)}}^{[\Delta \Delta]}_{}$
is real negative in the quasi-elastic region
(see Eq.~(\ref{eq:NC})).
Therefore $R^{[N\Delta]}_{L}$ is positive
but $R^{[N\Delta]}_{T}$ is negative.
This is the reason why
$R^{[NN]}_{L}$ and $R^{[N\Delta]}_{L}$
contribute constructively,
whereas
$R^{[NN]}_{T}$ and $R^{[N\Delta]}_{T}$
do destructively.

\subsection{Sum rules}
Next we consider the energy-non-weighted and
energy-weighted sums
defined by
\begin{eqnarray}
X^{0}_{L,T}(q) &=& \int R^{}_{L,T}(q,\omega) d\omega, \\
X^{1}_{L,T}(q)  &=& \int \omega R^{}_{L,T}(q,\omega)
d\omega.
\end{eqnarray}
When $\Delta$ is not included,
the former behaves as
\begin{eqnarray}
X^{0}_{L,T}(q) \rightarrow 1 \hskip 3em (q \rightarrow
\infty).
\end{eqnarray}
Here we used $\mbox{$ \widetilde{j}$}^{}_{} = j^{}_{}$ for
isovector currents
since the isospin of the ground state is assumed to be zero.

In the Fermi gas model there is the definite upper limit
of the integral $\omega_{\rm max}$.
In the case without $\Delta$, it is
\begin{eqnarray}
\omega_{\rm max}^{\rm F} = \frac{q^{2}}{2m} + \frac{qp_{\rm
F}}{m},
\end{eqnarray}
with the Fermi momentum $p_{\rm F}$.
For the finite nucleus
the upper limit extends to infinity in principle,
because there is no sharp cutoff of the momentum
distribution.
However the response functions is small and decrease
rapidly
beyond $\omega_{\rm max}^{\rm F}$.
Therefore
we assume exponential damping beyond $\omega_{\rm max}^{\rm
F}$
to evaluate the integrations.

In the case with $\Delta$,
$R^{}_{}(q,\omega)$ is sizable
both in
the $ph$ and
$\Delta h$ sectors.
For the sake of comparison, however,
we make the same prescription as
in the case without $\Delta$,
because we are interested in the strength distributed
only in the quasi-elastic region in this paper.

We present
the energy-non-weighted sums $X^{0}_{L,T}(q)$
in Fig.~\ref{fig:WFIG_M}(a) for $^{40}$Ca.
We took
$p_{F} = 1.20 \;{\rm fm}^{-1}$ to estimate $\omega^{\rm
F}_{\rm max}$.
The sums $X^{0}_{L}$ and $X^{0}_{T}$
of the
uncorrelated responses
are too close to distinguish.
We found that
they are also very close to the Fermi gas model value
\begin{eqnarray}
X^{0}_{\rm FG}(q)
   =
   \left\{\begin{array}{ll}
   \frac{3}{4} Q \left( 1 - \frac{Q^{2}}{12}\right) &
\hskip 2em {\rm for} \hskip 1em Q \leq 2 \\
   1                                                &
\hskip 2em {\rm for} \hskip 1em Q  >   2
          \end{array}
   \right.,
\end{eqnarray}
with $Q \equiv q / p_{\rm F}$
\cite{Forest66}.
For the uncorrelated responses reduction from unity
reflects the Pauli blocking effect.

The RPA correlation without $\Delta$
slightly increase $X^{0}_{L}$
but largely decreases $X^{0}_{T}$
as were seen in Fig.~\ref{fig:WFIG_A}(a).
Once $\Delta$ is included,
$X^{0}_{L}$ is drastically enhanced,
while $X^{0}_{T}$ is more quenched in the low $q$ region
but the quenching becomes smaller as $q$ increases.

In Fig.~\ref{fig:WFIG_M}(b),
the energy-weighted sums $X^{1}_{L,T}$ are shown.
We found that
the RPA correlation without $\Delta$
hardly changes them from the uncorrelated cases,
therefore
we did not show the results for this case.
Once $\Delta$ is introduced,
the sum of the spin-longitudinal mode $X^{1}_{L}$
is enhanced very much,
whereas that of the spin-transverse mode $X^{1}_{T}$
is only slightly affected.

The sum rule says that
the sums of the response function for an operator $\hat{O}$
are given by
\begin{eqnarray}
 X^{0}(q) &=& \langle\, \Psi_{0} \,
|\,\hat{O}\hat{O}^{\dag}\,|\, \Psi_{0} \, \rangle/A
 , \\
 X^{1}(q)  &=& \langle\, \Psi_{0} \, |\,[\hat{O}, [H,
\hat{O}]]\,|\, \Psi_{0} \, \rangle/A
 ,
 \label{Esum}
\end{eqnarray}
where $\hat{O}$ is assumed to be hermitian in
Eq.~(\ref{Esum}).
Pandharipande {\it et al.}
\cite{Pandharipande94}
calculated $\,|\, \Psi_{0} \, \rangle$ exactly
by using realistic nuclear force
within the nucleon degree of freedom.
Then they evaluated
$X^{0}$ and $X^{1}$
from these sum rules.
It is found that their sum rule values are
significantly larger than our results without $\Delta$.
Fig.~\ref{fig:WFIG_Q}(a) and (b) compare their values with
ours for $^{16}$O.
For instance,
their energy-non-weighted sums are
about 19 \% (longitudinal) and 10 \% (transverse)
and
their energy-weighted sums are
about 75 \% (longitudinal) and 55 \% (transverse)
larger than ours,
at $q = 1.70 \;{\rm fm}^{-1}$.
Such larger difference strongly indicates
importance of the correlations beyond RPA.
Note that in their calculation
the effects of $\Delta$ are implicitly included in part
through the $\Delta$ mediated three body force,
but the processes expressed by
$R^{[N\Delta]}_{}$ and
$R^{[\Delta\Delta]}_{}$
are not included.

Thouless
\cite{Thouless61}
proved
that one gets the energy-weighted sum of the RPA response
function
by replacing $\,|\, \Psi_{0} \, \rangle$ in Eq.~(\ref{Esum})
by $\,|\, \Psi^{HF}_{0} \, \rangle$, the Hartree-Fock
ground state wave function.
This theorem explains why
the RPA energy-weighted sums without $\Delta$
are very close to uncorrelated ones.
It also supports that
the difference
between our results and those of Pandharipande {\it et
al.}\
must be due to
the nuclear correlations beyond RPA.
This should also be reflected
in the energy spectra of the response functions.
The importance of the 2$p$-2$h$ configuration mixing
has been also pointed out by several authors
\cite{Alberico84,Takayanagi93}.

\section{Dependence on Effective Interactions}
\label{sec:EffectivInt}

In this section
we investigate the effective interaction dependence of
$R^{}_{L,T}$
and their process-decomposed components
$R^{[\alpha\beta]}_{L,T}$.
Some of the effective interactions are shown in
Fig.~\ref{fig:WFIG_O}
in the form of
$W^{[\alpha\beta]}_{L,T}/
(\frac{f_{\alpha}f_{\beta}}{m_{\pi}^{2}})$.
Their
$[\alpha\beta]$ dependence
comes only through $g'_{\alpha\beta}$.

We present
the $g'_{NN}$ dependence of
$R^{}_{L}$ and
$R^{}_{T}$
in Fig.~\ref{fig:WFIG_C}(a) and (b),
and
their $g'_{N\Delta}$ dependence
in Fig.~\ref{fig:WFIG_C}(c) and (d), respectively,
for $^{40}$Ca at $q = 1.70 \;{\rm fm}^{-1}$.
We fixed  $g'_{N\Delta}=0.4$ and $g'_{\Delta \Delta}=0.5$
in the study of $g'_{NN}$ dependence
and $g'_{NN}=0.6$ and $g'_{\Delta \Delta}=0.5$
for $g'_{N\Delta}$ dependence.
The response functions so weakly depend on
$g'_{\Delta\Delta}$
that we do not discuss about it.

Both $R^{}_{L}$ and $R^{}_{T}$ considerably depend on
$g'_{NN}$ as well as $g'_{N\Delta}$.
As $g'_{NN}$ decreases,
$R^{}_{L}$ becomes more enhanced at lower $\omega$
but less at higher $\omega$, while
$R^{}_{T}$ becomes less quenched at the lower side and
less enhanced at the higher side.
The $g'_{N\Delta}$ dependence is more simply summarized.
As $g'_{N\Delta}$ decreases,
both $R^{}_{L}$ and $R^{}_{T}$ increase,
namely,
$R^{}_{L}$ is more enhanced but
$R^{}_{T}$ less quenched.
These features are qualitatively common
for all nuclei $^{40}$Ca, $^{16}$O and $^{12}$C we
analyzed.
Such dependence is more clearly seen through
the process-decomposed response functions.
The $g'_{}$ dependence of
$R^{[\alpha\beta]}_{L,T}$
is shown in Fig.~\ref{fig:WFIG_J}.
Since $g'_{N\Delta}$ controls the coupling strength
between $N$ and $\Delta$,
$R^{[N\Delta]}_{}$ is more sensitive
to $g'_{N\Delta}$ than $g'_{NN}$,
whereas opposite is true for $R^{[NN]}_{}$.

Their behaviors are well understood
by the interpretation given at the end of
Subsec.~\ref{sec:DeltaEffect}.
As $g'_{NN}$ decreases
the effective interaction ${W_{L}}^{[NN]}_{}$ becomes
more attractive as shown in Fig.~\ref{fig:WFIG_O}.
Consequently
$R^{[NN]}_{L}$ is more enhanced at lower $\omega$
but more reduced at higher $\omega$.
On the other hand,
${W_{T}}^{[NN]}_{}$ becomes less repulsive and therefore
$R^{[NN]}_{T}$ is less quenched at lower $\omega$
but less enhanced at higher $\omega$.
These behaviors reflect in the $g'_{NN}$ dependence
of $R^{}_{L}$ and $R^{}_{T}$
seen in Fig.~\ref{fig:WFIG_C}.

As $g'_{N\Delta}$ decreases,
${W_{L}}^{[N\Delta]}_{}$ becomes more negative but
${W_{T}}^{[N\Delta]}_{}$ does less positive.
Therefore
both $R^{[N\Delta]}_{L}$ and $R^{[N\Delta]}_{T}$
are increased,
consequently more enhancement of $R^{}_{L}$
and less quenching of $R^{}_{T}$ are resulted in.

G.E. Brown and his collaborators
\cite{Brown91,Brown94}
advocated
the scaling of effective masses of
nucleon and mesons (except for pion) in the nucleus; {\it
e.g.}
$m^{\ast}_{\rho}/m_{\rho} \approx m^{\ast}_{N}/m_{N}$.
Correspondingly they claimed necessity of large $g'_{NN}$
\cite{Brown94}.
To see an implication of this proposal,
we show in Fig.~\ref{fig:WFIG_K}
the transverse response functions $R^{}_{T}$
with smaller $m^{\ast}_{\rho}(=0.75 m_{\rho})$ and
larger $g'_{NN}(= 0.8)$
together with the uncorrelated and the RPA responses
with $m^{\ast}_{\rho}=m_{\rho}$ and $g'_{NN} = 0.6$
as reference.
We did not change the nucleon effective mass
because our computer program cannot take into account
its density dependence at the present
and it must be the free nucleon mass at infinity.
So the present calculation aims only
to get feeling about the interaction dependence.

Let us compare two cases,
(a) a previous parameter set [$m^{\ast}_{\rho}=m_{\rho}$,
$(g'_{NN}, g'_{N\Delta}, g'_{\Delta\Delta})$
$=(0.6, 0.4, 0.5)$],
and
(b) the new one [$m^{\ast}_{\rho}=0.75 m_{\rho}$,
$(g'_{NN}, g'_{N\Delta}, g'_{\Delta\Delta})$
$= (0.8, 0.4, 0.5)$].
In Fig.~\ref{fig:WFIG_O}(b)
we also show ${W_{T}}^{}_{}$
with smaller $m^{\ast}_{\rho}(=0.75 m_{\rho})$.
It shows that
${W_{T}}^{[NN]}_{}$ is almost the same for the both cases
around $q \approx 1.70 \;{\rm fm}^{-1}$ accidentally,
and hence similar hardening is expected.
However
the positive ${W_{T}}^{[N\Delta]}_{}$ in the case (a)
becomes very weak negative in the case (b),
and hence $R^{[N\Delta]}_{T}$ changes the sign.
Consequently we see in Fig.~\ref{fig:WFIG_K} that
the hardening of $R^{}_{T}$ stays similar for the both
cases
but quenching is very much reduced in the latter.
We must note that
such change strongly depends on the momentum $q$
as is seen in
Figs.~\ref{fig:WFIG_O},
\ref{fig:WFIG_H} and
\ref{fig:WFIG_I}.
We remark that $R^{}_{L}$ is also affected
through the change of $g'_{NN}$.
Its enhancement is  reduced at lower $\omega$
because of large $g'_{NN}$.

Next
we show the collectivity ratio
$R^{}_{L}/R^{}_{T}$ in Fig.~\ref{fig:WFIG_D},
for
(a) $^{40}$Ca and
(b) $^{12}$C
at $q = 1.70\;{\rm fm}^{-1}$.
In the uncorrelated case,
the ratio is, of course, close to unity,
deviation from which is only due to
the single-particle spin-orbit force.
In the RPA calculation without $\Delta$,
the ratio is larger than unity at lower $\omega$
but smaller at higher $\omega$.
It is because
$R^{}_{L}$ is
enhanced at lower $\omega$ and
quenched at higher $\omega$,
but $R^{}_{T}$ behaves in the opposite way
as was shown in Fig.~\ref{fig:WFIG_A}(a).
In the case with $\Delta$,
the RPA correlation drastically increases the ratio
and
makes it larger than unity almost for whole $\omega$.
The case with
$(g'_{NN}, g'_{N\Delta}, g'_{\Delta\Delta}) = $
(0.6, 0.4, 0.5)
gives larger ratio at lower $\omega$
than the almost universality case with
(0.6, 0.6, 0.5)
and changes more steeply as $\omega$ changes.
Smaller $g'_{N\Delta}$ brings down
more $R^{}_{L}$ strength
from the $\Delta h$ region.
In the case with
$m^{\ast}_{\rho} = 0.75 m_{\rho}$,
enhancement of $R^{}_{L}$ is suppressed
due to larger $g'_{NN} = 0.8$ ,
thus the ratio is smaller than the above two cases.

The RPA effects became larger as the mass number increases.
We note that these results of $R^{}_{L}/R^{}_{T}$
cannot be compared with the $(p,n)$ data at the present
stage,
because effects of distortion and absorption
have not yet been considered.

\section{Electron Scattering}
\label{sec:ElectronScattering}

In this section
we study the transverse response functions
$R^{(e,e')}_{T}{}{}$
obtained by electron scattering (see Eq.~(\ref{eq:Ree})).

To analyze the electron scattering,
it has been known that we must take account of
not only RPA correlation with $\Delta$ but also the mixing
of
$2p$-$2h$ or more complicate configurations
\cite{Alberico84,Takayanagi93}
and the exchange currents
\cite{Alberico84,Carlson94},
etc..
Here we do not intend to reproduce the experimental data,
but want to see to what extent
the RPA results depend on
the magnetic transition ratio
$f_{\gamma N\Delta}/f^{IV}_{\gamma NN}$
and the effective interactions.

As was mentioned in Sec.~\ref{sec:Introduction},
most of previous analyses used the ratio
$f_{\gamma N\Delta}/f^{IV}_{\gamma NN} = 2.0$.
However, the SU(6) quark model
\cite{Close79}
gives
$f_{\gamma N\Delta}/f^{IV}_{\gamma NN}
= 6\sqrt{2}/5 \approx 1.70$.
Phenomenological analyses by Koch {\it et al.}
\cite{Koch84}
and Kumano
\cite{Kumano89}
gave the ratio 2.20 and 2.26, respectively.
A reason for the discrepancy is that
the $\pi N$ background scattering is
renormalized in the phenomenological analyses
but
should be treated explicitly in the quark model
\cite{Tanabe85}.

In Fig.~\ref{fig:WFIG_P},
we compare the results with different values of
$f_{\gamma N\Delta}/f^{IV}_{\gamma NN}$.
The smaller the ratio the smaller the quenching, because
the effect of $\Delta$ is essentially determined by the
product
$2 (f_{\gamma N\Delta}/f^{IV}_{\gamma NN})
R^{[N\Delta]}_{T}$.
We see some dependence on
$f_{\gamma N\Delta}/f^{IV}_{\gamma NN}$
when $R^{[N\Delta]}_{T}$ is sizable.

We compare the results with various effective interactions
for $^{12}$C  at $q = 300{\;\rm MeV}/c$ in
Fig.~\ref{fig:WFIG_H}(a)
and at $400{\;\rm MeV}/c$ in (b),
and
for $^{40}$Ca at $330{\;\rm MeV}/c$ in
Fig.~\ref{fig:WFIG_I}(a)
and at $410{\;\rm MeV}/c$ in (b).
Here we fixed
$f_{\gamma N\Delta}/f^{IV}_{\gamma NN} = 2.20$
\cite{Koch84}.
Experimental data are shown as a reference.
Compared with the energy spectrum of the uncorrelated case,
the experimental spectrum is very much hardened,
but the magnitudes are comparative.

The RPA result with
$(g'_{NN}, g'_{N\Delta}, g'_{\Delta\Delta})$
$ = (0.6, 0.6, 0.5)$,
which is practically the same as that of the universality
ansatz
(0.6, 0.6, 0.6),
is quenched and hardened.
As a result, the peak moves closer to the observed
position,
but the magnitude becomes much smaller than the data.

The calculation with $(0.6, 0.4, 0.5)$,
which we used as the standard in Sec.~\ref{sec:DeltaHole},
places the peak at slightly higher energy
and now at almost the right position.
It also increases the magnitude
though it is still smaller than the data.

If we take
$m^{\ast}_{\rho} = 0.75 m_{\rho}$ and
$g'_{NN} = 0.8$
as Brown and Rho
\cite{Brown91}
suggested,
the magnitude is increased very much
and the peak comes very close to the experimental data.
A good fit is seen in Fig.~\ref{fig:WFIG_H}(b)
but overshooting in Fig.~\ref{fig:WFIG_I}(b).
We must note that the good fit does not necessarily mean
that
the effective interaction is good
because there must be other contributions.

Qualitative features of our analysis are consistent
with previous calculation of Alberico {\it et al.}
\cite{Alberico86},
in which the RPA correlation gives reasonable hardening
but underestimates the magnitude,
the deficiency of which may be fulfilled
by nuclear correlations beyond RPA (the $2p$-$2h$ effects,
etc.)
and exchange currents.

Since $R^{(e,e')}_{T}$ eminently depends on the effective
interactions,
it must be a good tool to discriminate them
if the reliable estimation is possible of the other
contributions
such as $2p$-$2h$ configuration mixing and exchange
currents, etc..
Their estimation is beyond the scope of the present paper.

\section{Summary}
\label{sec:Summary}
We studied the effects of the $\Delta$-hole configurations
on the spin-isospin response functions
in finite nuclei in the quasi-elastic region.
We removed the universality ansatz
for the Landau-Migdal parameters and
treated $g'_{NN}$, $g'_{N\Delta}$ and
$g'_{\Delta\Delta}$ independently.
For this sake we formulated the response function method
for a finite system consisting of $N$ and $\Delta$
in the ring approximation.

We showed that the $\Delta$-mixing is crucial
for the enhancement of $R^{}_{L}$
in the whole range of the quasi-elastic region,
and it promotes
the quenching of $R^{}_{T}$.
If $\Delta$ is not included,
$R^{}_{L}$ and $R^{}_{T}$ are both
partially enhanced and partially quenched.
We emphasize that
reliable estimation of the effects of $\Delta$
is definitely needed
for comparison with experimental data.

Detailed analysis was carried out
by dividing the response functions $R^{}_{L,T}$
into the process-decomposed ones,
$R^{[NN]}_{L,T}$,
$R^{[N\Delta]}_{L,T}$ and
$R^{[\Delta\Delta]}_{L,T}$.
The main contribution comes from $R^{[NN]}_{L,T}$
but contribution from $R^{[N\Delta]}_{L,T}$
is also significant,
whereas $R^{[\Delta\Delta]}_{L,T}$ is negligible.
The effects of $\Delta$ is mostly represented
by $R^{[N\Delta]}_{L,T}$.

We showed that
$R^{[NN]}_{L}$ and $R^{[N\Delta]}_{L}$
contribute constructively,
whereas
$R^{[NN]}_{T}$ and $R^{[N\Delta]}_{T}$
do destructively.
This is the reflection
that the negative interaction $W^{[N\Delta]}_{L}$
between $ph$ and $\Delta h$
brings down the spin-longitudinal strength
from the $\Delta h$ to $ph$ region,
but the positive interaction $W^{[N\Delta]}_{T}$
brings up the spin-transverse strength in the opposite
direction.
Thus $R^{[\Delta N]}_{L}$ plays an important role
for strong enhancement of $R^{}_{L}$,
and
$R^{[\Delta N]}_{T}$ does some role
for quenching of $R^{}_{T}$.

Analysis of $g'_{}$ dependence of $R^{[\alpha\beta]}_{}$
tells that
$R^{[N\Delta]}_{}$ is very sensitive to $g'_{N\Delta}$
but not to $g'_{NN}$,
whereas
$R^{[NN]}_{}$ is sensitive to $g'_{NN}$
but not to $g'_{N\Delta}$.
As $g'_{N\Delta}$ decreases,
both $R^{}_{L}$ and $R^{}_{T}$ increase,
thus $R^{}_{L}$ is more enhanced
but $R^{}_{T}$ less quenched.
As $g'_{NN}$ decreases,
$R^{}_{L}$ becomes more enhanced at lower $\omega$ side
but reduced at higher side, while
$R^{}_{T}$ becomes less quenched at lower side
but more at the higher side.
Consequently
the choice of $g'_{NN}$ and $g'_{N\Delta}$ is crucial
to determine the behavior of the response functions
$R^{}_{L,T}$,
such as the collectivity ratio $R^{}_{L}/R^{}_{T}$.
Thus we should relax the universality condition for
$g'_{}$'s.
Effect of change of the $\rho$-meson effective mass
is also presented.

We further studied the interaction dependence of
the isovector transverse response functions
$R^{(e,e')}_{T}{}{}$
for $(e,e')$ scattering
in comparison with the experimental data.
The comparison must be useful
to investigate the effective interaction
at large $q$,
though we need reliable estimation of the exchange
currents,
2p-2h configuration mixing, etc..

In this paper
we present the detailed analysis for $^{40}$Ca.
The similar analysis for $^{12}$C is given
in Ref.~\cite{Ichimura94}.

At the end we itemize some of the remaining problems
we have to investigate.
1)
For analysis of data of hadronic probes
we must investigate reaction mechanisms
such as distortions, multistep processes, etc..
This will be discussed in the forth-coming paper.
2)
As we often mentioned
nuclear correlations beyond RPA have to be evaluated.
3)
We used Woods-Saxon shell model with the free nucleon mass.
Hartree-Fock field should be used to keep consistency
between the mean field and the residual interactions.
4)
Nucleon effective mass with position dependence
should be incorporated.
5)
Spreading widths of $ph$ propagation should be
properly taken into account.
At the present we only included that of the particles
by an energy independent complex potential.

\section*{Acknowledgments}

We would like to thank
Prof.~V.R.~Pandharipande for valuable discussions
and
Dr.~K.~Kawahigashi for useful advice on program coding.
This work is supported by the Grant-in-Aid
for Scientific Research of
the Ministry of Education (No.~02640215, 05640328).
The computer calculation for this work has been
financially supported in part
by Research Center for Nuclear Physics, Osaka University.
We used the terminal emulation software, {\em Eterm\/},
programmed by K.~Koketsu and K.~Takano,
Earthquake Research Institute, University of Tokyo,
and appreciate them.

\newpage
\section*{Figure Captions}
\begin{figure}[h]

\caption{\label{fig:WDIAG_A}
Process-decomposed response functions.
}
\end{figure}

\begin{figure}[h]

\caption{\label{fig:WFIG_A}
Isovector spin-response functions,
$R^{}_{L}$ and $R^{}_{T}$,
for $^{40}$Ca at $q = 1.70 \;{\rm fm}^{-1}$.
(a) Without and
(b) with $\Delta$.
The dotted and full lines denote $R^{}_{L}$ and $R^{}_{T}$
with RPA correlation, respectively.
The dot-dashed and dashed lines represent
$R^{(0)}_{L}$ and $R^{(0)}_{T}$, respectively.
$(g'_{NN}, g'_{N\Delta}, g'_{\Delta\Delta}) = (0.6, 0.4,
0.5)$
are used in RPA.
}
\end{figure}

\begin{figure}[h]

\caption{\label{fig:WFIG_F}
$R^{}_{L}$ and $R^{}_{T}$ for
(a) $^{16}$O and
(b) $^{12}$C
at $q = 1.70 \;{\rm fm}^{-1}$ with $\Delta$.
The notations of the lines and
the values of $g'_{}$
are same as those in Fig.~{\protect\ref{fig:WFIG_A}}.
}
\end{figure}

\begin{figure}[h]

\caption{\label{fig:WFIG_B}
Process-decomposed response functions,
(a) $R^{[\alpha\beta]}_{L}$ and
(b) $R^{[\alpha\beta]}_{T}$,
for $^{40}$Ca at $q = 1.70 \;{\rm fm}^{-1}$.
The dotted, dashed and dot-dashed lines denote
$R^{[NN]}_{L,T}$,
$2 (f_{\Delta}/f_{N}) R^{[N\Delta]}_{L,T}$, and
$(f_{\Delta}/f_{N})^{2} R^{[\Delta\Delta]}_{L,T}$,
respectively,
appeared in the Eq.~({\protect\ref{eq:RLRTtotal}}).
$R^{}_{L}$ and
$R^{}_{T}$
are shown by the full lines.
The values of $g'_{}$
are same as those in Fig.~{\protect\ref{fig:WFIG_A}}.
}
\end{figure}

\begin{figure}[h]

\caption{\label{fig:WFIG_N}
Response functions
for $^{40}$Ca at $q = 1.70 \;{\rm fm}^{-1}$
without $\Delta$ calculated in TDA
($g'_{NN} = 0.6$).
The dotted and full lines show $R^{}_{L}$ and $R^{}_{T}$,
respectively.
The dot-dashed and dashed lines represent
$R^{(0)}_{L}$ and $R^{(0)}_{T}$, respectively.
}
\end{figure}

\begin{figure}[h]

\caption{\label{fig:WFIG_M}
(a) Energy-non-weighted sums $X^{0}$ and
(b) energy-weighted sums $X^{1}$
for $^{40}$Ca.
The full line denotes
the sums of the uncorrelated transverse response.
The dotted line represents
the RPA transverse results without $\Delta$.
The RPA transverse and longitudinal ones with $\Delta$
are shown by
the dashed and dot-dashed lines, respectively.
The uncorrelated longitudinal sum
and
the RPA longitudinal sum without $\Delta$ are
very close to the uncorrelated transverse sum and not
shown.
}
\end{figure}

\begin{figure}[h]

\caption{\label{fig:WFIG_Q}
(a) Energy-non-weighted sums $X^{0}$ and
(b) energy-weighted sums $X^{1}$
for $^{16}$O.
The dashed and dot-dashed lines denote
the transverse and longitudinal sums, respectively,
calculated by Pandharipande {\it et al.}
{\protect\cite{Pandharipande94}}.
Our results without $\Delta$ are shown by
the full (transverse) and dotted (longitudinal) lines.
}
\end{figure}

\begin{figure}[h]

\caption{\label{fig:WFIG_O}
Effective interactions
$W^{[\alpha\beta]}_{L,T} /
(\frac{f_{\alpha}f_{\beta}}{m_{\pi}^{2}})$
at $\omega = 80 {\;\rm MeV}$;
(a) longitudinal and (b) transverse part.
At $q = 0 \;{\rm fm}^{-1}$
their values become equal to their $g'_{\alpha \beta}$.
The interactions
with $m_{\rho}^{\ast} = m_{\rho}$
and
with $m_{\rho}^{\ast} = 0.75 m_{\rho}$
are denoted by the full and dotted lines, respectively.
}
\end{figure}

\begin{figure}[h]

\caption{\label{fig:WFIG_C}
The $g'_{}$ dependence of the response functions
for $^{40}$Ca at $q = 1.70 \;{\rm fm}^{-1}$.
Left side:
the $g'_{NN}$ dependence of
(a) $R^{}_{L}$ and (b) $R^{}_{T}$
with $g'_{N\Delta}=0.4$ and $g'_{\Delta \Delta}=0.5$.
$g'_{NN}$ is set to be
0.5 (dotted line),
0.6 (dashed line) and
0.7 (dot-dashed line).
Right side:
the $g'_{N\Delta}$ dependence of
(c) $R^{}_{L}$ and (d) $R^{}_{T}$
with $g'_{NN}=0.6$ and $g'_{\Delta \Delta}=0.5$.
$g'_{N\Delta}$ is set to be
0.4 (dotted line),
0.5 (dashed line) and
0.6 (dot-dashed line).
The full line denotes the uncorrelated responses.
}
\end{figure}

\begin{figure}[h]

\caption{\label{fig:WFIG_J}
The $g'_{}$ dependence of
the process-decomposed response functions
for $^{40}$Ca at $q = 1.70 \;{\rm fm}^{-1}$.
The $g'_{NN}$ dependence of
(a) $R^{[\alpha \beta]}_{L}$ and (b) $R^{[\alpha
\beta]}_{T}$,
and
the $g'_{N\Delta}$ dependence of
(c) $R^{[\alpha \beta]}_{L}$ and (d) $R^{[\alpha
\beta]}_{T}$.
The notations of lines and
the values of $g'_{}$
are same as those in Fig.~{\protect\ref{fig:WFIG_C}}.
The uncorrelated responses are not shown here.
}
\end{figure}

\begin{figure}[h]

\caption{\label{fig:WFIG_K}
Effective $\rho$-meson mass
dependence of the transverse response function
for $^{40}$Ca at $q = 1.70 \;{\rm fm}^{-1}$
with the large $g'_{NN}$.
The dot-dashed line represents
the RPA result with
$m^{\ast}_{\rho} = 0.75 m_{\rho}$
and
$(g'_{NN}, g'_{N\Delta}, g'_{\Delta\Delta}) = (0.8, 0.4,
0.5)$.
As reference,
the uncorrelated response (full line) and
the RPA one (dotted line)
with
$(g'_{NN}, g'_{N\Delta}, g'_{\Delta\Delta}) = (0.6, 0.4,
0.5)$
are also shown.
}
\end{figure}

\begin{figure}[h]

\caption{\label{fig:WFIG_D}
Collectivity ratio
$R^{}_{L}/R^{}_{T}$ for
(a) $^{40}$Ca and
(b) $^{12}$C
at $q = 1.70\;{\rm fm}^{-1}$.
The thin full line denotes
the ratio without the correlation.
The dotted line represents
the RPA results without $\Delta$
[$g'_{NN}$ = 0.6].
The RPA results with $\Delta$ are shown by
the dashed line
[$(g'_{NN}, g'_{N\Delta}, g'_{\Delta\Delta}) = $
(0.6, 0.6, 0.5)],
the thick full line
[$(g'_{NN}, g'_{N\Delta}, g'_{\Delta\Delta}) = $
(0.6, 0.4, 0.5)],
and
the dot-dashed line
[$m^{\ast}_{\rho} = 0.75 m_{\rho}$
and
$(g'_{NN}, g'_{N\Delta}, g'_{\Delta\Delta}) = $
(0.8, 0.4, 0.5)],
respectively.
}
\end{figure}

\begin{figure}[h]

\caption{\label{fig:WFIG_P}
Dependence on $f_{\gamma N\Delta}/f^{IV}_{\gamma NN}$
of $S_{T}$
for $^{12}$C$(e,e')$ at $q = 300{\;\rm MeV}/c$.
$f_{\gamma N\Delta}/f^{IV}_{\gamma NN}$ is set to be
1.70 (dotted line),
2.00 (dashed line) and
2.20 (full line).
$(g'_{NN},g'_{N\Delta},g'_{\Delta\Delta})$ is set to be
(0.6, 0.6, 0.5).
}
\end{figure}

\begin{figure}[h]

\caption{\label{fig:WFIG_H}
$S_{T}$ for $^{12}$C$(e, e')$
with various effective interactions
at
(a) $q = 300 {\;\rm MeV}/c$ and
(b) $400 {\;\rm MeV}/c$.
$f_{\gamma N\Delta}/f^{IV}_{\gamma NN}$ is set to be 2.20.
The dotted and dashed lines denote
the RPA results
with
$(g'_{NN},g'_{N\Delta},g'_{\Delta\Delta}) =$
(0.6, 0.4, 0.5) and
(0.6, 0.6, 0.5), respectively.
The dot-dashed line represents
the RPA result
with
$(g'_{NN}, g'_{N\Delta}, g'_{\Delta\Delta}) =$
(0.8, 0.4, 0.5) and
$m^{\ast}_{\rho} = 0.75 m_{\rho}$.
The uncorrelated one is shown by the full line.
The experimental data are taken from
Ref.~{\protect\cite{Barreau83}}.
}
\end{figure}

\begin{figure}[h]

\caption{\label{fig:WFIG_I}
$S_{T}$ for $^{40}$Ca$(e, e')$
with various effective interactions
at
(a) $q = 330 {\;\rm MeV}/c$ and
(b) $410 {\;\rm MeV}/c$.
The notations of the lines are same as those in
Fig.~{\protect\ref{fig:WFIG_H}}.
The experimental data are taken from
Ref.~{\protect\cite{Alberico86}}.
}
\end{figure}

\end{document}